\documentclass[a4paper,fleqn,usenatbib]{mnras}

\usepackage{newtxtext,newtxmath} 

\usepackage[T1]{fontenc}
\usepackage{ae,aecompl}

\usepackage{bm}
\usepackage{graphicx}	
\usepackage{amsmath}	
\usepackage{amssymb}	
\usepackage{listings}
\usepackage{physics}

\newcommand{\program}[1]{{\scriptsize {\MakeUppercase{#1}}}}

\title[MAGI]{MAGI: many-component galaxy initialiser}

\author[Y. Miki and M. Umemura]{
Yohei Miki$^{1}$\thanks{E-mail: ymiki@cc.u-tokyo.ac.jp (YM)}
and Masayuki Umemura$^{2,3}$
\\
$^{1}$Information Technology Center, University of Tokyo, 5-1-5 Kashiwanoha, Kashiwa, Chiba 277-8589, Japan\\
$^{2}$Center for Computational Sciences, University of Tsukuba, 1-1-1 Tennodai, Tsukuba, Ibaraki 305-8577, Japan\\
$^{3}$CREST, JST, 1-1-1 Tennodai, Tsukuba, Ibaraki 305-8577, Japan
}

\date{Accepted XXX. Received YYY; in original form ZZZ}

\pubyear{2017}

\begin{document}
\label{firstpage}
\pagerange{\pageref{firstpage}--\pageref{lastpage}}
\maketitle

\begin{abstract}
Providing initial conditions is an essential procedure for numerical simulations of galaxies. 
The initial conditions for idealised individual galaxies in $N$-body simulations should resemble observed galaxies and be dynamically stable for time scales much longer than their characteristic dynamical times. 
However, generating a galaxy model ab initio as a system in dynamical equilibrium is a difficult task, since a galaxy contains several components, including a bulge, disc, and halo. 
Moreover, it is desirable that the initial-condition generator be fast and easy to use. 
We have now developed an initial-condition generator for galactic $N$-body simulations that satisfies these requirements. 
The developed generator adopts a distribution-function-based method, and it supports various kinds of density models, including custom-tabulated inputs and the presence of more than one disc. 
We tested the dynamical stability of systems generated by our code, representing early- and late-type galaxies, with $N=$~2,097,152 and 8,388,608 particles, respectively, and we found that the model galaxies maintain their initial distributions for at least 1~Gyr. 
The execution times required to generate the two models were $8.5$ and $221.7$ seconds, respectively, which is negligible compared to typical execution times for $N$-body simulations. 
The code is provided as open-source software and is publicly and freely available at \url{https://bitbucket.org/ymiki/magi}. 
\end{abstract}

\begin{keywords}
Galaxies: general -- Galaxies: structure -- Galaxies: bulges -- Galaxies: halos -- Galaxies: evolution -- Methods: numerical
\end{keywords}



\section{Introduction}
\label{section:introduction}
$N$-body simulations are powerful tools for investigating the dynamical evolution of galaxies including, e.g., galactic mergers or the development of spiral arms. 
In order to run $N$-body simulations, it is important to set up appropriate initial conditions. 
The initial system of $N$-body particles should be in dynamical equilibrium for at least the longest dynamical timescale in the problem. 
For example, the longest dynamical timescale may be the crossing time of a satellite galaxy in a galactic minor merger, or it may be the rotation time of a galactic disc in a simulation to follow the development of spiral arms. 
However, producing a system in dynamical equilibrium that represents a galaxy is difficult, since galaxies, in general, consist of several components, e.g., bulge, disc, and halo. 
The construction of initial conditions for $N$-body simulations is an on-going research topic, and many earlier studies have tackled this tough problem \citep{Hernquist1993, KuijkenDubinski1995, Boily2001, Widrow2003, WidrowDubinski2005, McMillanDehnen2007, Widrow2008, Vasiliev2013, Perret2014, YurinSpringel2014, VasilievAthanassoula2015}. 

A model system of particles that represents an observed galaxy is useful not only in providing initial conditions for $N$-body simulations, but also in creating mock observations to fit observed datasets. 
Recent progress with observatories like Gaia \citep{GaiaDR1_astrometry} demands ever-better models to study galactic dynamics in much greater detail. 
Since Gaia provides observational data in a six-dimensional phase space, direct comparisons between particle models and the Gaia data can provide information about the phase-space distribution function (hereafter, DF) of the Milky Way galaxy. 

Various density-profile models have been proposed and employed to cover the diversity of simulated and observed galaxies. 
For example, the Navarro--Frenk--White (hereafter, NFW) profile \citep{Navarro1995, Navarro1996}, the Moore profile \citep{FukushigeMakino1997, Moore1998}, and the Einasto profile \citep{Einasto1965, Navarro2004, Navarro2010} are often used to represent the density profiles for dark matter halos in cosmological $N$-body simulations with $\Lambda$ cold dark matter. 
For the bulge component, the well-known de Vaucouleurs's law \citep{deVaucouleurs1948} and the S\'ersic profile \citep{Sersic1963, CiottiBertin1999}--which is a generalisation of de Vaucouleurs's law--are frequently employed to fit the observed surface-density profiles. 
The Hernquist profile \citep{Hernquist1990}, which resembles de Vaucouleurs's law, has an analytic DF. 
For the disc component, an exponential disc or a S\'ersic profile are often used to fit the surface-density profiles of observed disc galaxies and to investigate their dynamical evolution in numerical studies. 
The presence of more than one disc component in a galaxy may be a general property of late-type galaxies. 
The Milky Way has both a thin and a thick disc \citep{Juric2008}, and observations by \citet{DalcantonBernstein2002, YoachimDalcanton2006} have revealed multiple disc components (thin and thick discs) in most edge-on, late-type galaxies. 

Many earlier studies have been dedicated to developing initial condition generators for $N$-body simulations; \citet{Hernquist1993} pioneered this research field. 
The approach by \citet{Hernquist1993} approximates the velocity distribution by a local Maxwellian and calculates the velocity dispersion using the Jeans equation. 
\citet{Boily2001} generalised the approach for axisymmetric bulges and halos. 
\citet{Springel2005} also extended the work by \citet{Hernquist1993} to include a gaseous disc, and \citet{Perret2014} provided implementation as the open-source software \program{DICE}. 
However, \citet{Kazantzidis2004} showed that particle systems based on the local Maxwellian approximation are not always in dynamic equilibrium, which may lead to inadequate interpretations of results from $N$-body simulations that use such initial conditions. 
Currently, the standard approach is to employ DF-based implementations. 
A widely used example in the community is the software \program{GalactICS} \citep{KuijkenDubinski1995, Widrow2003}. 
This code generates a dynamically stable system that employs a lowered Evans model \citep{KuijkenDubinski1994} for the halo, a King model \citep{Michie1963, MichieBodenheimer1963, King1966} for the bulge, and an exponential disc. 
Alternatively, one can adopt orbit-based methods like the Schwarzschild method or an iterative method by \citet{Rodionov2009} to generate self-consistent dynamical-equilibrium systems of particles. 
Removing assumptions for DFs--e.g., spherical symmetry of the system and isotropy of velocity distribution--extends the scope of applications to non-axisymmetric systems and makes the codes more flexible than DF-based implementations. 
However, the execution time is much longer than for DF-based methods. 
\citet{VasilievAthanassoula2015} reported that the execution time required to produce an axisymmetric disc galaxy model with $10^6$ particles reached an hour or much longer with both the \program{GALIC} code \citep{YurinSpringel2014} and the \program{SMILE} code \citep{Vasiliev2013, VasilievAthanassoula2015}, while \program{GalactICS} took only a few minutes. 
In this study, we have accordingly adopted a DF-based method rather than an orbit-based method to reduce the time-to-solution. 

To generate a model galaxy that resembles an observed galaxy, the initial-condition generator needs to be flexible. 
The physical properties of the components represented by the $N$-body particles must be easily adjustable in order to facilitate investigations of the dependence of physical processes on the mass, size, or profile of a galaxy. 
For example, the particle system generated by \program{GalactICS} is dynamically stable and is useful for numerical investigations of galactic dynamics. 
However, changing the particle system to represent accurately an observed galaxy is cumbersome. 
Some input parameters in \program{GalactICS}--e.g., the potential and the velocity dispersion at the centre--depend on the density profile of the system, making it difficult for the user to control basic quantities such as the total mass of the system. 
Furthermore, \program{GalactICS} internally solves Poisson's equation and modifies the mass distribution toward a dynamically stable system. 
Therefore, a trial-and-error procedure is necessary to generate the desired particle system. 
Changing the input parameters from kinematic ones (such as the value of potential and the velocity dispersion) to structural ones (such as the total mass and scale length), and fixing the mass distributions of the spherical components, would alleviate the burden on the user performing these iterations. 

The following are the key requirements we set for our initial-condition generator: (1) dynamical stability of the particle system generated, (2) ability to represent various kinds of density profiles, (3) ability to produce more than one disc component, and (4) simplicity and convenience of use. 
We have realised these requirements in our initial-condition generator for $N$-body simulations, called \program{MAGI} (MAny-component Galaxy Initialiser). 
The generator adopts a DF-based method, and it supports various kinds of density-model inputs, including a machine-readable tabular format and the presence of multiple disc components. 
The code is provided as open-source software. 
The organisation of this manuscript is as follows: 
Section~\ref{section:methods} introduces methods and assumptions adopted in \program{MAGI}. 
The stability of the particle systems generated is tested in Section~\ref{section:results}. 
In Section~\ref{section:discussion}, we examine the validity of some numerical treatments and measure the execution time of \program{MAGI}. 
Section~\ref{section:conclusions} summarises our work. 

\section{Methods}
\label{section:methods}
In this section, we describe the DF-based implementation of our initial-condition generator, \program{MAGI}.
Sections~\ref{subsec:methods:eddington} and \ref{subsec:methods:abel} describe how to calculate the DF and discuss the further techniques required to generate a spherical component for a given surface-density profile. 
One of the strong points of \program{MAGI} is a high degree of freedom in generating disc components. 
We describe the theories, assumptions and numerical techniques employed for the disc components in Section~\ref{subsec:methods:disc}. 
Finally, Section~\ref{subsec:methods:implementation} contains other implementation details. 

\subsection{Eddington's Formula}
\label{subsec:methods:eddington}
In this section, we derive the DF for a case in which the system is spherically symmetric. 
Consider a system consisting of $N$ components with a volume-density profile $\rho_i(r)$, where the subscript $i$ specifies the $i$-th component. 
If we assume an isotropic velocity distribution, Eddington's formula \citep{Eddington1916, BinneyTremaine2008} gives the DF $f_i$ of the $i$-th component as 
\begin{equation}
  f_i(\mathcal{E})
  = \frac{1}{\sqrt{8}\pi^2} \left[\int_0^{\mathcal{E}} \dd \Psi\, \frac{1}{\sqrt{\mathcal{E} - \Psi}} \dv[2]{\rho_i}{\Psi} + \frac{1}{\sqrt{\mathcal{E}}} \left.\dv{\rho_i}{\Psi}\right|_{\Psi=0}\right], 
  \label{eq:methods:eddington:DF}
\end{equation}
\begin{equation}
  \mathrm{where}\,\,\Psi = \sum_{i = 1}^N \Psi_i, 
\end{equation}
and where $\mathcal{E}$ and $\Psi$ are the relative energy per unit mass and the relative potential of the system, respectively. 
The last term in equation~(\ref{eq:methods:eddington:DF}) corresponds to the density gradient at infinity, which vanishes. 
Now the second derivative of the mass density with respect to the relative potential can be rewritten as 
\begin{equation}
  \dv[2]{\rho_i}{\Psi}
  = \left(\frac{r^2}{G M(r)}\right)^2 \left[\dv[2]{\rho_i}{r} + \dv{\rho_i}{r} \left(\frac{2}{r} - \frac{4 \pi r^2 \sum_i \rho_i(r)}{M(r)}\right)\right],
  \label{eq:methods:eddington:derivative}
\end{equation}
which is much easier to implement \citep{Kazantzidis2006} compared to equation~(\ref{eq:methods:eddington:DF}). 
We have calculated the DF using equations~(\ref{eq:methods:eddington:DF}) and (\ref{eq:methods:eddington:derivative}) for cases in which the density profile and its first and second derivatives are given for the individual components. 
Since $M(r)$ is the total mass enclosed within radius $r$, the mass distributions of all components, including the disc components and the mass of the central massive black hole, must be given before calculating the DF. 
The mass profiles of the disc components are treated as spherically averaged profiles of the enclosed mass (i.e., the zeroth order multipole approximation). 

\program{MAGI} provides various density profiles for generating models commonly adopted in astrophysics. 
The functional forms of the volume-density profiles supported by \program{MAGI}, together with their first and second derivatives with respect to radius, are summarised in Appendix~\ref{appendix:rho}. 
The supported density profiles that have a central core are the Plummer profile \citep{Plummer1911}, the Burkert profile \citep{Burkert1995}, the King profile (the profile and derivatives are determined numerically using equations given in Appendix~\ref{appendix:King}), and the King profile which is given in empirical form \citep{King1962}. 
Density profiles that have a central cusp are also implemented: the Hernquist profile, the NFW profile, the Moore profile, and the Einasto profile. 
\program{MAGI} also supports two broken-power-law density profiles, the double-power-law model \citep{Hernquist1990, Merritt2006} given by equation~(\ref{eq:methods:eddington:double}) and the triple-power-law model described by equation~(\ref{eq:methods:eddington:triple}). 
If the system contains a central massive black hole, the black hole particle is placed at the centre-of-mass of the system with zero velocity. 

\program{MAGI} also accepts density profiles given in machine-readable tabular form. 
Since evaluating the first and second derivatives of the profile by differencing the tabulated values can result in significant loss of accuracy, especially when the profile has a central core, \program{MAGI} interpolates the profile using a cubic-spline curve, which is then used to compute tables of the derivatives. 
Once the density profile and its derivatives are tabulated from the input density profile, the subsequent steps are identical to the case for a predefined density profile. 

To generate systems of particles in dynamical equilibrium, in some cases we need to specify an explicit cutoff radius $r_\mathrm{c}$. 
If a cutoff radius is specified, \program{MAGI} multiplies the profiles with a complementary-error-function-based smoother of the form
\begin{equation}
  \frac{1}{2} \mathrm{erfc}\left(\frac{r - r_\mathrm{c}}{2 \varDelta_\mathrm{c}}\right) = 
  \frac{1}{2} \left(1 - \frac{2}{\sqrt{\pi}} \int_0^{(r - r_\mathrm{c}) / 2 \varDelta_\mathrm{c}} \dd t \, e^{-t^2}\right), 
\end{equation}
where $\varDelta_\mathrm{c}$ is the smoothing scale. 

\subsection{Abel Transformation}
\label{subsec:methods:abel}
In most cases, observations provide surface-density profiles rather than the volume-density profiles discussed in Section~\ref{subsec:methods:eddington}. 
\program{MAGI} also supports the input of a surface-density profile. 
De-projection from the input surface-density profile $\Sigma(R)$ to a volume-density profile $\rho(r)$ is performed by an Abel transformation \citep[cf.,][]{BinneyTremaine2008}: 
\begin{equation}
  \rho(r) = -\frac{1}{\pi} \int_r^\infty \dd R \, \dv{\Sigma}{R} \frac{1}{\sqrt{R^2 - r^2}}, 
  \label{eq:methods:abel:transformation}
\end{equation}
if $\rho(r)$ drops faster than $r^{-1}$. 
\program{MAGI} constructs the volume-density profile as a numerical table by applying the Abel transformation numerically. 
The derivatives of the density profile are given by cubic-spline interpolation, as in the case for which the input volume-density profile is a machine-readable table. 
Once the density profile and its derivatives are obtained in tabular form from the input surface-density profile, the subsequent steps are the same as in the case for which the input profile is a volume-density profile. 
\program{MAGI} also supports the S\'ersic profile described in Appendix~\ref{appendix:Sigma}, which includes de Vaucouleurs's law, and the surface-density profiles for this case are provided as machine-readable tables. 

\subsection{Disc Components}
\label{subsec:methods:disc}
The disc is one of the most important characteristics of a late-type galaxy. 
Observations by \citet{DalcantonBernstein2002} and by \citet{YoachimDalcanton2006} established that most edge-on, late-type galaxies have multiple disc components (both thin and thick discs). 
Our initial-condition generator must therefore be capable of generating several disc components in dynamical equilibrium. 
\program{MAGI} is the first initial-condition generator to support more than one disc components. 

We have extended the approach adopted in \program{GalactICS} \citep{KuijkenDubinski1995, Widrow2003} to provide particle distributions that include multiple disc components. 
In \program{GalactICS}, the disc component is assumed to have an exponential surface-density profile, with an isothermal profile in the vertical direction. 
The potential-density pair is derived by solving Poisson's equation using a spherical multipole expansion with a simple approximation for high-order terms. 
\program{MAGI} supports S\'ersic profiles (the exponential profile corresponds to a S\'ersic profile with S\'ersic index unity), and surface-density profiles are specified in machine-readable tabular form for the disc components. 
Supporting S\'ersic profiles having various S\'ersic indices is essential for representing a large variety of observed disc components. 
For example, \citet{Kelvin2012} reported that a typical S\'ersic index is around $0.5-2$ for disc components obtained from the GAMA (Galaxy And Mass Assembly) database. 
We have also assumed that every disc component has an isothermal profile in the vertical direction, given by 
\begin{equation}
  \rho(R, z) \propto \Sigma(R) \exp{\left(-\frac{\Phi(R, z) - \Phi(R, 0)}{\Phi(R, z_\mathrm{d}) - \Phi(R, 0)}\right)},
\end{equation}
where $z_\mathrm{d}$ is the scale height of the disc component. 
The vertical profile of a disc component thus represents the density field as a function of the potential field, while Poisson's equation gives the potential field as a function of the density field. 
Therefore, we derive the potential-density pair for the superposition of all components numerically by iterating the following two procedures until convergence is obtained: (1) Determine the vertical profile of the individual disc components from the potential field, and (2) solve Poisson's equation to determine the potential field from the superposition of the density fields of all components. 
We solve Poisson's equation using the BiCGSTAB method preconditioned with ILU(0) \citep{vanderVorst1992, Itoh2012}. 
We use a nested grid to discretize the density and potential fields, and the number of levels--typically ten--is automatically determined. 
The outer boundary condition at $R = R_\mathrm{max}$ or $z = z_\mathrm{max}$ for level $L = 0$ (the coarsest grid) is derived from \citet{BinneyTremaine2008} by assuming a constant disc-scale-height. 
The outer boundary condition for level $L > 0$ is set by the potential field in the coarser level $L - 1$. 
For the inner boundary conditions ($R = 0$ or $z = 0$), $\Phi(\pm R, \pm z) = \Phi(R, z)$ is set from the symmetry about $R = 0$ and $z = 0$. 
The number of grid points needed to determine the potential-density pair is 256 in the $R$ direction and 64 in the $z$ direction, respectively. 
We adopted the full multigrid method proposed by \citet{PressTeukolsky1991}. 
In order to determine the velocity structure, we adopted the Schwarzschild DF \citep{Schwarzschild1907, BinneyTremaine2008}: 
\begin{equation}
  f_\mathrm{Sch}(L_z, z) \propto \exp{\left[-\frac{{v_R}^2 + \gamma^2 \left\{v_\phi - v_c(R_\mathrm{g})\right\}^2}{2 {\sigma_R}^2(L_z)} - \frac{{v_z}^2}{2 {\sigma_z}^2(L_z)}\right]},
\end{equation}
where $L_z$ is the $z$-component of the specific angular momentum, and $v_R$, $v_\phi$, and $v_z$ are the velocities in the horizontal, azimuthal, and vertical directions, respectively. 
Also, $R_\mathrm{g}$ is the guiding-centre radius, $v_c(R) = R \Omega$ is the circular velocity, where $\Omega$ is the circular frequency, and $\sigma_R$ and $\sigma_z$ are, respectively, the velocity dispersions in the horizontal and vertical directions. 
The remaining quantity, $\gamma$, is defined as $2 \Omega_\mathrm{g} / \kappa$, where $\Omega_\mathrm{g}$ is the circular frequency at the guiding-centre radius, and $\kappa$ is the epicycle frequency. 
The underlying assumption behind the Schwarzschild DF is the epicycle approximation, assuming small velocity dispersion. 

The radial profile of the velocity dispersion in the horizontal direction, $\sigma_R(R)$, is another problem. 
\program{GalactICS} assumes an exponential profile 
\begin{equation}
  \sigma_R^2(R) \propto \exp{(- R / R_\mathrm{s})},
  \label{eq:methods:disc:galactics}
\end{equation}
where $R_\mathrm{s}$ is the disc scale length, even though \program{GalactICS} adopts the epicycle approximation, assuming that the orbit is nearly circular. 
In \program{MAGI}, the velocity dispersion $\sigma_p$ in the azimuthal direction is equal to $\min{(\sigma_z, f v_c)}$, where $f$ is a scaling parameter. 
Then $\sigma_R(R)$ becomes 
\begin{equation}
  \sigma_R(R) = \gamma \sigma_p 
  \label{eq:methods:disc:magi}
\end{equation}
under the epicycle approximation. 
The velocity dispersion is thus smaller than that in \program{GalactICS}, especially in the central region, ensuring consistency with the epicycle approximation. 

In \program{GalactICS}, the circular velocity at the particle position is determined by the derivative of the potential in the equatorial plane of the disc component; in other words, the rotation velocities of the particles do not depend on $z$, while the potential does. 
This results in too fast a rotation to maintain the initial distribution, especially in a thick disc at large $z$, and a relaxation phase is required before dynamical stability is reached. 
\program{MAGI} improves on this by calculating the circular velocity from the gradient of the potential at the particle location. 

Finally, in order to generate a realistic distribution, we need a recipe for suppressing a needle-like structure that occasionally forms near the rotation axis of the disc. 
When the scale height $z_\mathrm{d}$ of the disc is not sufficiently small, compared to the scale length $r_\mathrm{sph}$ of the spherical components, the density decreases gradually in the vertical direction for $R \lesssim r_\mathrm{sph}$. 
This produces a needle-like structure near the rotation axis of the disc component (see Fig.~\ref{fig:discussion:needle}b); it is an artefact resulting from the isothermal profile assumed in the vertical direction. 
We have introduced a new quantity $h = h(R)$ which satisfies 
\begin{equation}
  \exp{\left(-\frac{\Phi\left(R, \min{\left(16 z_\mathrm{d}, r_\mathrm{c}\right)}\right) - \Phi(R, 0)}{\Phi(R, h(R)) - \Phi(R, 0)}\right)} = e^{-16}, 
  \label{eq:methods:disc:removal}
\end{equation}
and we replace the scale height $z_\mathrm{d}$ by $z_\mathrm{d}(R) = \min{(z_\mathrm{d}, h(R))}$. 
This ensures that the vertical density profile vanishes for $|z| \geq \min{(16 z_\mathrm{d}, r_\mathrm{c})}$ and damps the needle-like structure. 

\subsection{Configuration and Implementation}
\label{subsec:methods:implementation}
In this subsection we provide some details about the structure and use of \program{MAGI}. 
As pointed out in Section~\ref{section:introduction}, input parameters such as the mass and scale length are preferable to kinematic ones such as the central potential or velocity dispersion. 
In \program{MAGI}, the input parameter sets for each component are limited to the mass $M$, the scale length $r_\mathrm{s}$, and other dimensionless parameters that are specific to the adopted models, such as the S\'ersic index $n$ or the dimensionless King parameter $W_0$. 

\begin{figure}
\begin{lstlisting}[label=list:input, caption=Format of the input file.]
1 # select the system of units
4 # number of components
4	halo.param	0	0
1	bulge.param	0	0
1000	bh.param	1	1
-1	disc.param	0	0
\end{lstlisting}
\end{figure}
\begin{figure}
\begin{lstlisting}[label=list:bulge, caption=Format of the parameter file (King model).]
5.0e+10  # mass of the component
1.0      # scale radius of the component
5.0      # model specific parameter(s)
0        # set cutoff (1) or not (0)
\end{lstlisting}
\end{figure}
\begin{figure}
\begin{lstlisting}[label=list:disc, caption=Format of the parameter file (S\'ersic disc).]
2.5e+10  # mass of the component
5.0      # scale radius of the component
2.0      # Sersic index (only for Sersic disc)
1.0      # scale height of the component
-1.0 0.1 # parameters for velocity dispersion
0.0      # retrograde fraction [0.0, 1.0]
1        # set cutoff (1) or not (0)
50.0 5.0 # cut off radius and width
\end{lstlisting}
\end{figure}
Listings~\ref{list:input}--\ref{list:disc} are samples of the input files that show the expected format and parameters. 
The text following each ``\#'' symbol are brief comments describing the parameters; they do not appear in the actual input files. 
In Listing~\ref{list:input}, the first two integers specify the system of units to be used in the simulation and the number of components. 
The configuration of each component follows: the index of the model, the corresponding parameter file, whether to specify the number of particles (1) or not (0), and the specified number of particles. 
For the central massive black hole, the number of particles must be unity. 
Also, the configurations of the disc components must come after those of the spherical components. 

In each parameter file, the mass $M$ and scale radius $r_\mathrm{s}$ of the components appear first. 
If the unit ID $= 1$ (as specified in Listing~\ref{list:input}), the units of mass and length are $M_\odot$ and kpc, respectively. 
Since \program{MAGI} internally converts the system of units from astrophysical units (like $M_\odot$, kpc, and km~s$^{-1}$) to code units, users need not perform unit conversions. 
For the spherical components, model-specific parameters (if they exist) follow; for example, the non-dimensional parameter $W_0$ at the centre of the King model (Listing~\ref{list:bulge}). 
In the case of the disc components (e.g., Listing~\ref{list:disc}), the scale height $z_\mathrm{d}$ is an additional key parameter. 
The following two parameters pertain to the velocity-dispersion profile. 
The first sets the radial velocity dispersion at the centre, $\sigma_{R, 0}$, and the second sets the scaling parameter $f$. 
A negative value of $\sigma_{R, 0}$ indicates that the value is identical to the vertical velocity-dispersion at the centre. 
The first parameter is used only when the radial velocity-dispersion model given by equation~(\ref{eq:methods:disc:galactics}), which is that used in \program{GalactICS}, is adopted. 
By default, \program{MAGI} adopts the radial velocity-dispersion model given by equation~(\ref{eq:methods:disc:magi}) in the previous subsection. 
The next parameter is optional and allows the introduction of a retrograde component in the disc. 
The remaining parameters are related to the explicit cutoff of the density distribution. 
An integer indicates whether an explicit cutoff of the density profile is required (1) or not (0). 
When a cutoff is set (specified as 1), the cutoff radius $r_\mathrm{c}$ and its smoothing scale $\varDelta_\mathrm{c}$ are passed to the software by specifying them in the next line. 

\program{MAGI} employs the inverse-function method and the rejection method, respectively, to determine the particle positions and velocities. 
Pseudo-random numbers are generated by a SIMD (single instruction, multiple data)-oriented Fast Mersenne Twister of period $2^{19937} - 1$ \citep[\program{SFMT}\footnote{\url{http://www.math.sci.hiroshima-u.ac.jp/~m-mat/MT/SFMT/}} 1.5.1 by][]{SaitoMatsumoto2008} with a jump function\footnote{\url{http://www.math.sci.hiroshima-u.ac.jp/~m-mat/MT/SFMT/JUMP/index.html}}. 
The jump function guarantees the independence of multiple series of random numbers generated in parallel, while just changing the initial values of a random number generator cannot strictly ensure independence. 
In the current version of \program{MAGI}, the program is parallelized using OpenMP and shifts the initial state for \program{SFMT} in each OpenMP thread by $10^{100}$ steps. 
The particles' positions and velocities are shifted component-by-component to set the centre-of-mass of the system at the origin of the coordinate system and remove the bulk motion. 

\program{MAGI} uses Gaussian quadrature for some numerical integrations, employing GSL\footnote{\url{https://www.gnu.org/software/gsl/}} 2.4 for this purpose. 
The output format of the particle data is a structure of arrays contained in the HDF5\footnote{\url{https://support.hdfgroup.org/HDF5/}} 1.8.18 container or the TIPSY format\footnote{\url{https://github.com/N-BodyShop/tipsy/wiki/Tipsy}}, depending on the user's preference. 
The default configuration is the HDF5 container. 

\section{Results}
\label{section:results}
This section examines the dynamical stability of particle distributions generated by \program{MAGI}. 
Section~\ref{subsec:results:etg} considers the stability of a spherically symmetric system like an early-type galaxy or the bulge of disc galaxy. 
Section~\ref{subsec:results:ltg} studies the stability of an axisymmetric system having two disc components with different scale heights. 
The initial condition is generated on a workstation (Intel Xeon E5-2640v3, 16 cores, 2.60 GHz, DDR4-2133, 64GB, gcc 4.8.5). 
The code is compiled with the following options for performance optimisation: \texttt{-O3 -ffast-math -funroll-loops -march=native -fopenmp}. 
The code is run via the \texttt{numactl} command with the \texttt{---localalloc} option. 
Time evolution is calculated using the octree code \program{GOTHIC} \citep{MikiUmemura2017}, which runs on NVIDIA GeForce GTX TITAN X with nvcc 8.0.44. 

\subsection{Stability of a Spherically Symmetric System}
\label{subsec:results:etg}
\begin{figure}
  \includegraphics[viewport=0 0 409 683, width=\columnwidth, clip]{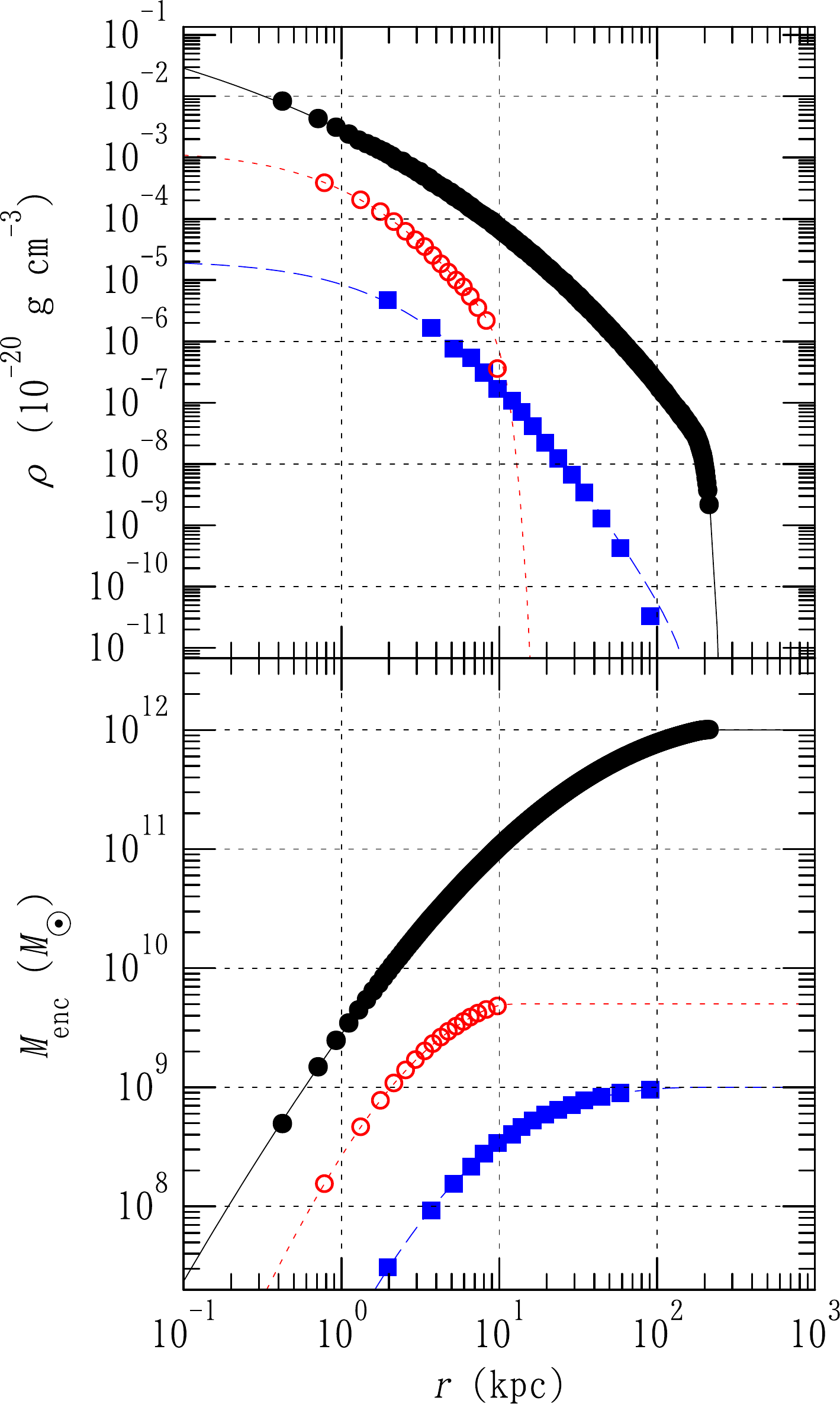}
  \caption{
    Volume-density profile (\textit{upper panel}) and enclosed-mass profile (\textit{lower panel}) of a model galaxy that represents an early-type galaxy with $2^{21} =$~2,097,152 particles. 
    Symbols and curves show the profile at $t = 1$~Gyr and the input profile, respectively. 
    Every pair of symbols and curves represents different components of the model galaxy: the dark-matter halo with an Einasto profile (the black filled circles and the solid curve), the stellar halo with a triple-power-law profile (the blue filled squares and the dashed curve), and the bulge with a profile obeying the de Vaucouleurs's law (the red open circles and the dotted curve). 
  }
  \label{fig:galaxy:spherical.density.profile}
\end{figure}
We first benchmark \program{MAGI} with a spherically symmetric model in order to confirm the effectiveness of the Eddington formula. 
We chose the mass ratios of the spherical components in our model from well-established observed correlations. 
The Magorrian relation \citep{Magorrian1998, MarconiHunt2003} gives the correspondence between the galactic bulge and central massive black hole (MBH), where the BH mass is around $0.2$\% of that of the spheroidal component. 
A similar observational relation between the dark-matter halo and the central MBH \citep{Ferrarese2002} implies that the mass of the central black hole is around $10^7 M_\odot$, if the dark-matter halo mass is $10^{12} M_\odot$. 
In addition to the typical components of early-type galaxies, we added a stellar halo resembling that of the Milky Way in the model galaxy. 
Recent observations of the stellar halo of the Milky Way suggest that its power-law index is around $-3$ at $r \lesssim 50$~kpc and $-5$ in the outer halo \citep{Keller2008, Akhter2012}. 
The masses of the stellar halos of nearby disc galaxies with masses comparable to that of the Milky Way are $1-6 \times 10^9 M_\odot$ \citep{Harmsen2017}. 
In summary, the model galaxy is a superposition of an Einasto halo ($M = 10^{12} M_\odot$, $r_\mathrm{s} = 10$~kpc, $\alpha = 0.2$, $r_\mathrm{c} = 200$~kpc, $\varDelta_\mathrm{c} = 10$~kpc) with 2,084,644 particles; a stellar halo given as a triple-power-law model ($M = 10^9 M_\odot$, $r_\mathrm{in} = 3$~kpc, $r_\mathrm{out} = 50$~kpc, $\alpha = 0$, $\beta = 1$, $\gamma = 3$, $\delta = 1$, $\epsilon = 5$, $r_\mathrm{c} = 150$~kpc, $\varDelta_\mathrm{c} = 20$~kpc) with 2,084 particles; a stellar component obeying the de Vaucouleurs's law ($M = 5 \times 10^{9} M_\odot$, $r_\mathrm{s} = 2$~kpc, $r_\mathrm{c} = 10$~kpc, $\varDelta_\mathrm{c} = 1$~kpc) with 10,423 particles; and an MBH particle with a mass of $10^7 M_\odot$. 
The elapsed time for \program{MAGI} to generate this model galaxy was $8.5$ seconds. 

We have calculated the time evolution for this model galaxy over $1$~Gyr using \program{GOTHIC} with an accuracy-controlling parameter $\varDelta_\mathrm{acc} = 2^{-7} = 7.8125 \times 10^{-3}$ and a Plummer softening length of $15.625$~pc. 
The resulting radial profile is shown in Fig.~\ref{fig:galaxy:spherical.density.profile}. 
The figure compares the volume-density profile and the enclosed-mass profile for each component at $t = 1$~Gyr (symbols) with the input radial profiles (curves). 
The figure clearly shows that the given density profile is stable over the entire domain after an integration time of $1$~Gyr, which is approximately 70 times the free-fall time at $r = 2$~kpc. 
In other words, \program{MAGI} successfully generates the model galaxy as a system in dynamical equilibrium. 

\subsection{Stability of Multiple Disc Components}
\label{subsec:results:ltg}
\begin{figure}
  \includegraphics[viewport=0 0 759 716, width=\columnwidth, clip]{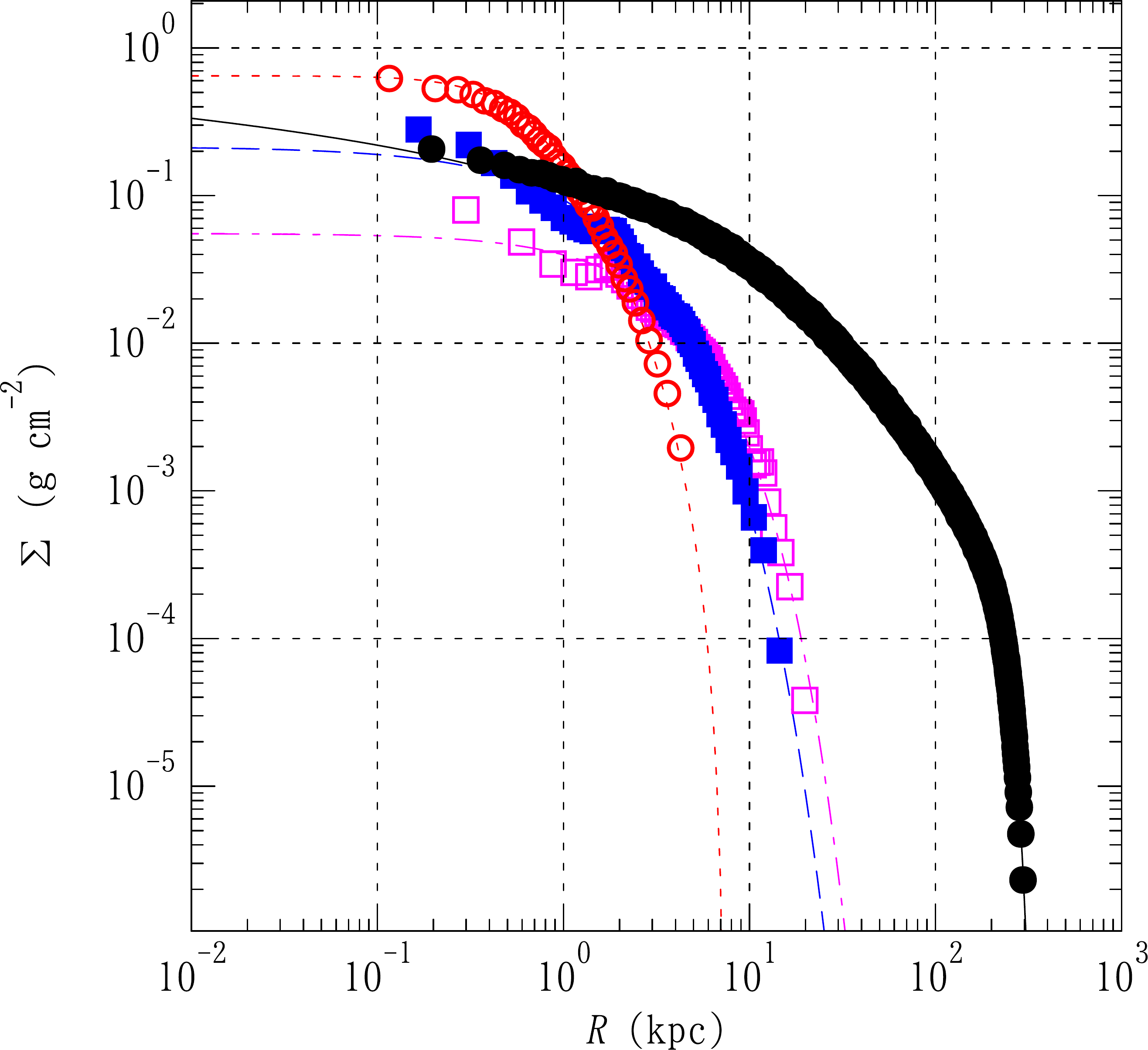}
  \caption{
    Surface-density profile of a model galaxy that uses $2^{23} =$~8,388,608 particles to represent a late-type galaxy. 
    Symbols and curves show the profile at $t = 1$~Gyr and the input profile, respectively. 
    Each pair of symbols and curves represents a different component of the model galaxy: the dark-matter halo represented with an NFW profile (the black filled circles and the solid curve), the bulge as a King model (the red open circles and the dotted curve), the thick disc with a S\'ersic profile (the blue filled squares and the dashed curve), and the thin disc as an exponential profile (the magenta open squares and the dot-dashed curve). 
  }
  \label{fig:ltg:surface.density.profile}
\end{figure}
\begin{figure}
  \includegraphics[width=\columnwidth, clip]{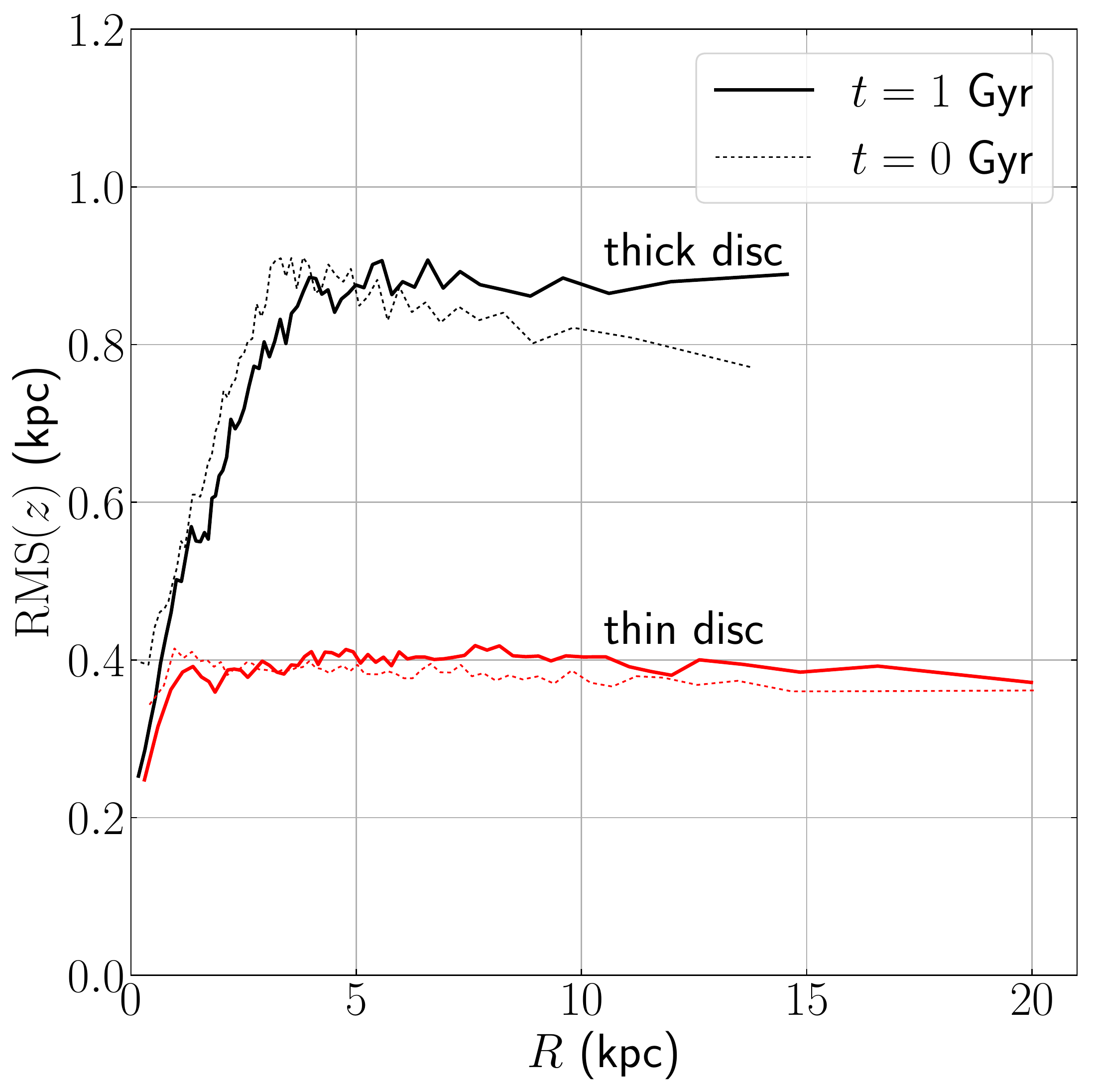}
  \caption{
    Radial profiles of disc thicknesses evaluated as the root-mean-square of disc particle heights about the $z = 0$ plane. 
    The black lines show the thick-disc component, while the red lines correspond to the thin-disc component. 
    The bold and regular lines represent $t = 1$~Gyr and the initial condition, respectively. 
  }
  \label{fig:ltg:disk.height}
\end{figure}
The next test involves models of late-type galaxies. 
In contrast to early-type galaxies, late-type galaxies often harbor multiple discs \citep{DalcantonBernstein2002, YoachimDalcanton2006}. 
We constructed a late-type galaxy model with an NFW halo ($M = 10^{12} M_\odot$, $r_\mathrm{s} = 20$~kpc, $r_\mathrm{c} = 250$~kpc, $\varDelta_\mathrm{c} = 20$~kpc) containing 8,065,971 particles; a King bulge ($M = 10^{10} M_\odot$, $r_\mathrm{s} = 0.7$~kpc, $W_0 = 5$) with 80,659 particles; a thick S\'ersic disc ($M = 1.5 \times 10^{10} M_\odot$, $R_\mathrm{s} = 3$~kpc, $n = 1.2$, $z_\mathrm{d} = 1$~kpc, $f = 0.125$) with 120,989 particles; and a thin exponential disc ($M = 1.5 \times 10^{10} M_\odot$, $R_\mathrm{s} = 3$~kpc, $z_\mathrm{d} = 0.5$~kpc, $f = 0.125$) with 120,989 particles. 
Toomre's $Q$-values for the thick and the thin discs at $R_\mathrm{s}$ are, respectively, $1.6$ and $2.2$. 
The bulge-to-total ratio of the model galaxy is $0.25$, which is consistent with observed late-type galaxies \citep{Oohama2009}. 
Other physical properties of the model galaxy are shown in Appendix~\ref{appendix:Disc}. 
The elapsed time required for \program{MAGI} to generate this model galaxy was $221.7$ seconds. 

We calculated the time evolution of this model galaxy over $1$~Gyr using \program{GOTHIC}, with an accuracy-controlling parameter $\varDelta_\mathrm{acc} = 2^{-8} = 3.90625 \times 10^{-3}$ and a Plummer softening length of $15.625$~pc. 
Figure~\ref{fig:ltg:surface.density.profile} compares the surface-density profile of each component at $t = 1$~Gyr (symbols) with the corresponding input profile (curves). 
The figure shows very satisfactory agreement between the imposed surface-density profile and the surface-density profile at $t = 1$~Gyr, which is around 220 times the free-fall time for the bulge component and around 10 times the rotation periods of the discs at their scale radii. 
Maintaining the respective disc thicknesses is an essential requirement for the disc components. 
Figure~\ref{fig:ltg:disk.height} compares the disc heights at $t = 1$~Gyr (the bold lines) with those set in the initial conditions (the regular lines), for the thin and thick discs separately (red and black lines, respectively). 
The disc heights are evaluated as the root-mean-square of the disc particle heights with respect to the $z = 0$ plane. 
Figure~\ref{fig:ltg:disk.height} shows that the thicknesses of both discs are almost unchanged after $1$~Gyr (around ten-fold the rotation time at $R = R_\mathrm{s}$). 
The drop in the thickness of the thick-disc component toward the centre is due to the scale-height reduction prescribed by equation~(\ref{eq:methods:disc:removal}) to remove the artificial needle-like structure (see also Fig.~\ref{fig:discussion:needle}). 
In a series of convergence tests, it has turned out that simulation runs with a larger number of disc particles, $N_\mathrm{d}$, result in almost identical distributions. 
On the other hand, runs with a smaller number of halo particles, $N_\mathrm{h}$, significantly thicken the disc thicknesses. 
We found that $N_\mathrm{h} \gtrsim 3 \times 10^6$ and $N_\mathrm{d} \gtrsim 10^5$ is a sufficient condition to stabilize the disc components in a viewpoint of reducing disc heating by halo particles. 
It is noteworthy that we were able to use the output from \program{MAGI} directly as the initial condition for the $N$-body simulation. 
In other words, for these tests, \program{MAGI} did not require any relaxation procedure before starting the simulations. 

\section{Discussion}
\label{section:discussion}

\subsection{Accuracy of MAGI}
\label{subsec:discussion:accuracy}
\begin{figure*}
  \includegraphics[width=.8\linewidth, clip]{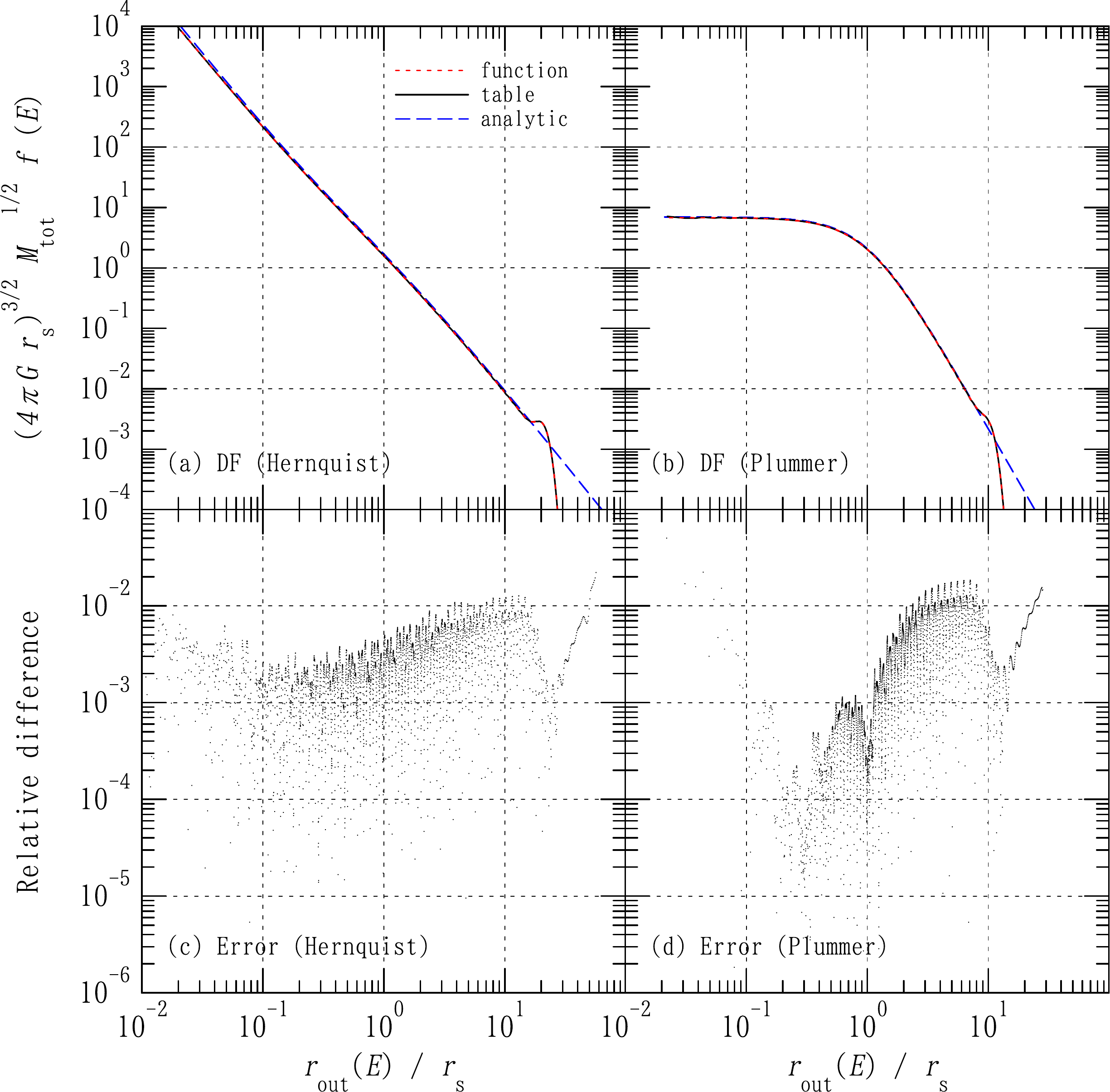}
  \caption{
    Comparison of DFs. 
    \textit{Upper panel}: DFs generated by \program{MAGI} using density profiles in functional form (the red dotted curves) or in machine-readable tabular form (the black solid curves) are shown, together with analytic expressions for the DFs (the blue dashed curves). 
    \textit{Lower panel}: The relative errors in the DFs based on tabular forms, as compared with the functional forms, are plotted as functions of the normalised apoapsis at the given energy. 
    The left panels present the results for the Hernquist profile, while the right panels correspond to the Plummer profile. 
  }
  \label{fig:discussion:df}
\end{figure*}
\begin{table}
  \centering
  \caption{
    Parameters of the compared density profiles. 
  }
  \label{tab:discussion:df}
  \begin{tabular}{ccccc}
    \hline
    Model & $M_\mathrm{tot}$~($M_\odot$) & $r_\mathrm{s}$~(kpc) & $r_\mathrm{c}$~(kpc) & $\varDelta_\mathrm{c}$~(kpc)\\
    \hline
    Hernquist & $10^{10}$ & 1 & 20 & 2\\
    Plummer   & $10^{9 }$ & 2 & 20 & 2\\
    \hline
  \end{tabular}
\end{table}
To validate the accuracy of the numerically generated DFs, we compared the DFs generated by \program{MAGI} with those given by analytic formulas. 
Figure~\ref{fig:discussion:df} shows the DFs for Hernquist and Plummer spheres, together with the differences originating from the input data format for \program{MAGI}. 
Here $r_\mathrm{out}(E)$ are the radii satisfying the relation 
\begin{equation}
  \Phi(r = r_\mathrm{out}) = E, 
\end{equation}
which corresponds to the apocentre for a given specific energy $E$. 
Table~\ref{tab:discussion:df} lists the physical properties of the profiles. 
The tables of density profiles have 128 bins equally spaced in the logarithm from $10^{-5} r_\mathrm{s}$ to $50 r_\mathrm{s}$. 
Figure~\ref{fig:discussion:df} indicates that the DFs derived from the density profiles in machine-readable tabular format match well with those given by the functional form. 
The relative error is below $\sim$2\% in most regions, irrespective of whether the central density profile is a cusp or a core. 
That is to say, \program{MAGI} properly generates DFs for the density profile given in machine-readable tabular form. 
Figs.~\ref{fig:discussion:df}(a) and \ref{fig:discussion:df}(b) also compare the DFs from \program{MAGI} with analytic expressions (equations \ref{eq:app:Hernquist:DF} and \ref{eq:app:Plummer:DF}). 
Since the DFs in \program{MAGI} have a density cutoff at a finite radius, they drop sharply around $r_\mathrm{c}$, while the analytic counterparts decrease continuously. 
The agreement between the DFs, except near the cutoff radius, confirms that the DF generator in \program{MAGI} works properly and with sufficient accuracy. 

\begin{figure}
  \includegraphics[width=\columnwidth, clip]{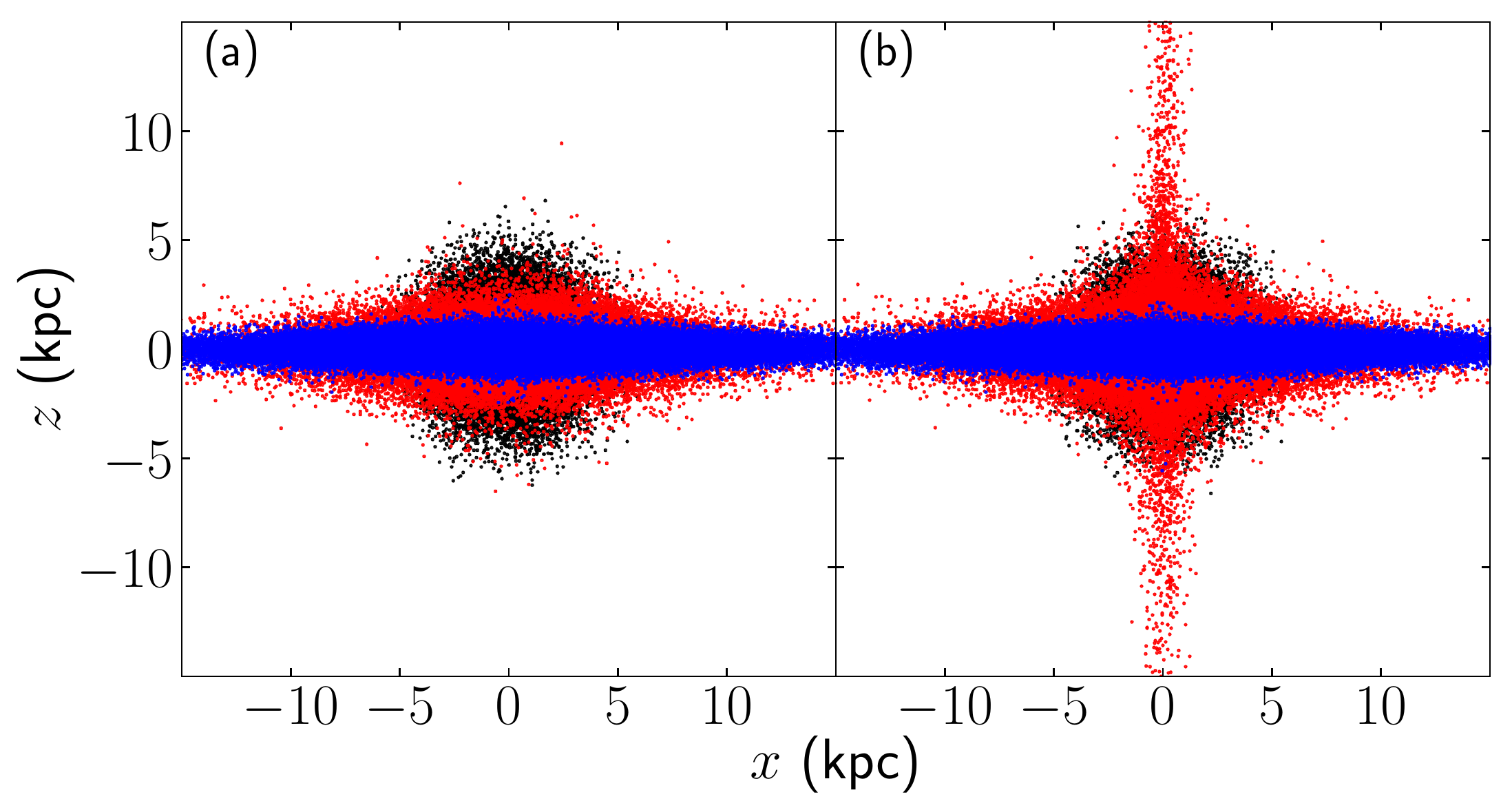}
  \caption{
    Removal of the artificial needle-like structure. 
    The black, red, and blue dots show the distributions of the $N$-body particles representing the bulge, thick-disc, and thin-disc components, respectively. 
    The left panel shows the particle distributions with the modification given by equation~(\ref{eq:methods:disc:removal}) and the right panel the distributions without any modification. 
  }
  \label{fig:discussion:needle}
\end{figure}
Figure~\ref{fig:discussion:needle} confirms that the additional treatment using equation~(\ref{eq:methods:disc:removal}) removes the artificial needle-like structure protruding from the disc (Fig.~\ref{fig:discussion:needle}a).
As shown in Fig.~\ref{fig:discussion:needle}(b), the needle-like structure appears only in the thick disc and not in the thin disc. 
The presence of the needle-like structure is a natural result of the assumption of an isothermal profile in the vertical direction. 
Because the thickness of the thick-disc component is comparable to the scale length of the bulge component, the vertical density profile near the rotation axis drops very slowly. 

\begin{figure*}
  \includegraphics[width=.9\linewidth, clip]{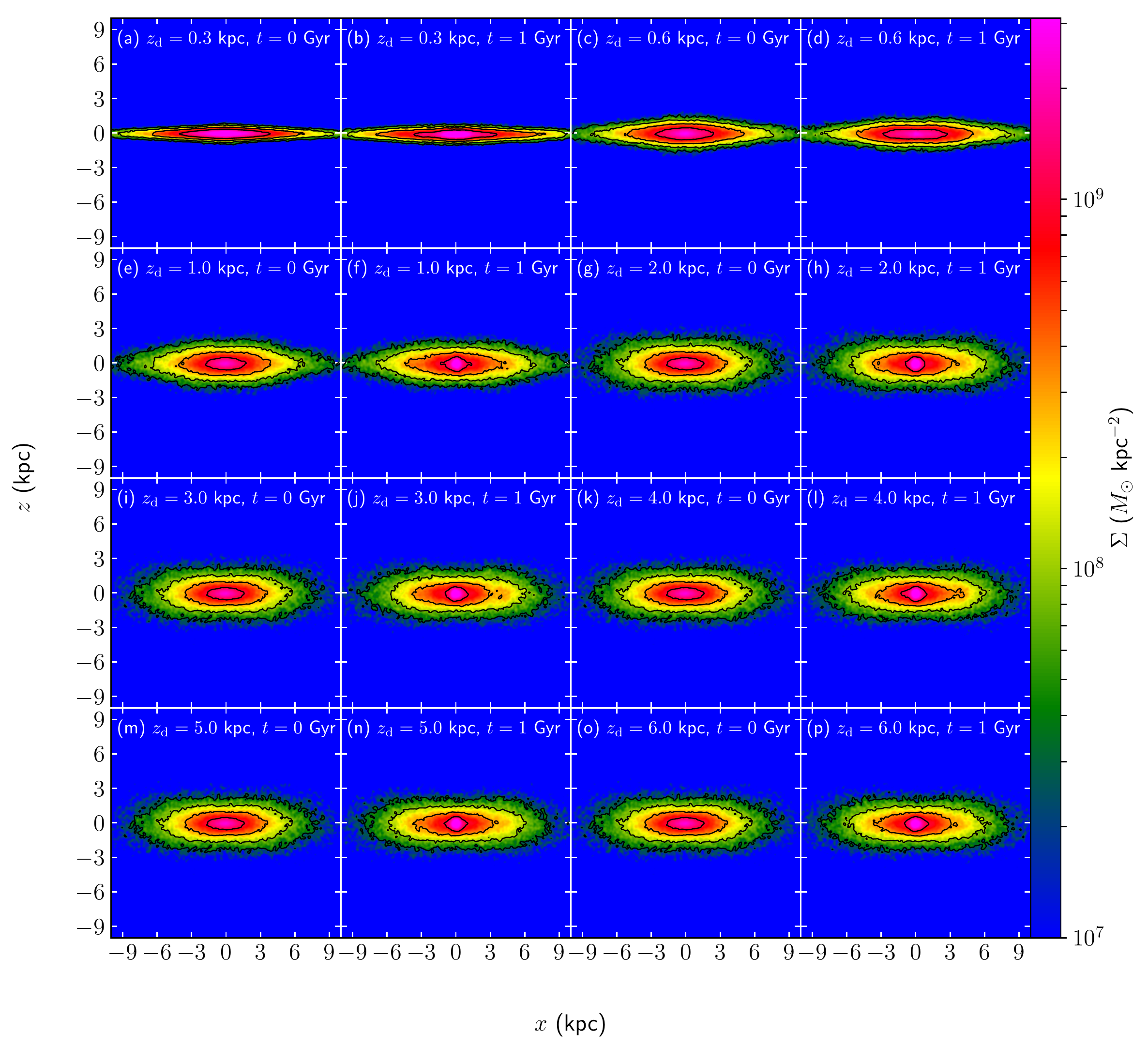}
  \caption{
    Surface-density distribution maps of thick discs at $t = 0$~Gyr and $1$~Gyr. 
    The model galaxies are identical to the late-type galaxy model described in Section~\ref{subsec:results:ltg}, except for the scale height of the thick disc, which is varied from $0.3$~kpc to $6$~kpc. 
  }
  \label{fig:discussion:scale.height}
\end{figure*}
The upper limit to the scale heights of disc components that work properly in \program{MAGI} is another concern. 
Since the extension to thick discs implemented in \program{MAGI} relies partly on the workaround described in Section~\ref{subsec:methods:disc}, there are limitations in the present method. 
To examine these limits, we have tested the dynamical stability of the thick-disc component by varying its scale height in the late-type galaxy model described in Section~\ref{subsec:results:ltg} from $R_\mathrm{s} / 10 = 0.3$~kpc to $2 R_\mathrm{s} = 6$~kpc. 
Figure~\ref{fig:discussion:scale.height} displays surface-density distribution maps of thick discs with various scale heights at $t = 0$~Gyr and $1$~Gyr. 
The figure shows that the surface-density distribution is stable for $1$~Gyr, even when the scale height exceeds the scale length $R_\mathrm{s} = 3$~kpc. 
This indicates that the thick disc is stable, at least for $z_\mathrm{d} \lesssim 2 R_\mathrm{s}$. 
We note that the morphology of the thick disc is similar at $z_\mathrm{d} \gtrsim R_\mathrm{s} = 3$~kpc because the scale height of the disc in the central region was reduced to $r_\mathrm{c} / 16 \sim 3$~kpc to remove the needle-like structure. 
If the removal of this artefact is incomplete, then infalling components from the needle-like structure generate a shell structure after having been scattered by the bulge. 
The masses of the needle-like structure and the resulting shell structure are negligibly small; however, numerical artefacts should be explicitly removed by capping the disc scale height according to equation~(\ref{eq:methods:disc:removal}). 

\subsection{Execution Time}
\label{subsec:discussion:time}
\begin{figure}
  \includegraphics[width=\columnwidth, clip]{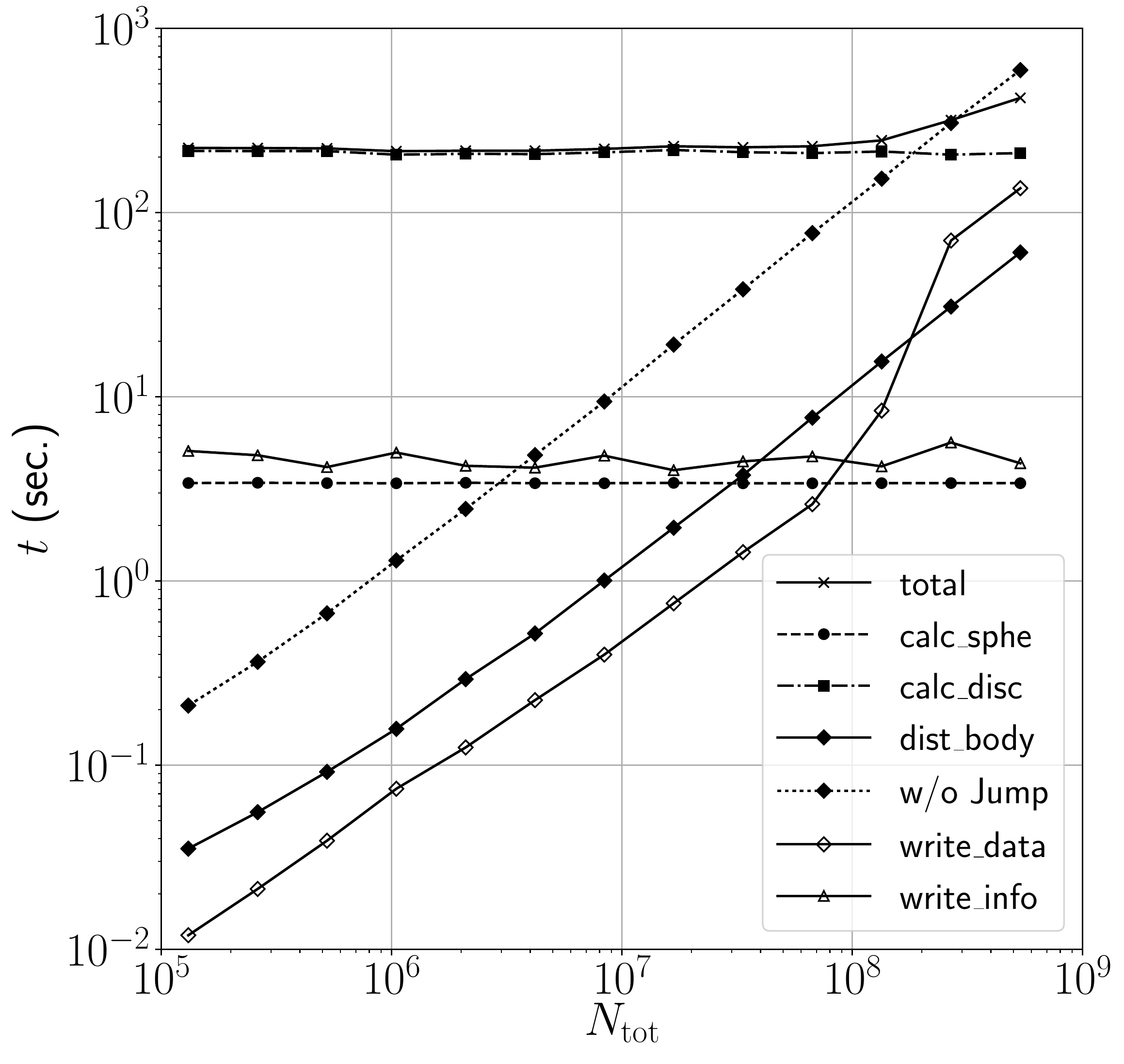}
  \caption{
    Execution time for the code as a function of the total number of $N$-body particles. 
    The crosses and solid line show the total running time for the code (with the use of the jump function), while the other markers and lines show the breakdown by different stages of the calculation. 
    The filled and open diamonds and solid lines are the times required to distribute the $N$-body particles according to the prescribed DF and the time required to dump the output file, respectively. 
    The filled diamonds and dotted line also show the execution time required to distribute the $N$-body particles, but for the case without the jump function. 
    The circles and dashed line, and the squares and dot-dashed line, respectively, represent the time necessary to calculate the DF for the spherical components and the disc components. 
    The triangles and solid line correspond to the time required to dump the files (the DF, density distribution, and various profiles such as the enclosed mass and potential). 
  }
  \label{fig:discussion:execution.time}
\end{figure}
The execution time of the software is an important performance metric. 
Figure~\ref{fig:discussion:execution.time} plots the medians of 10 measurements of the execution time required for \program{MAGI} to generate the late-type galaxy model tested in Section~\ref{subsec:results:ltg}. 
The problem size--that is, the number of $N$-body particles $N_\mathrm{tot}$--is varied from $2^{17} =$~131,072 to $2^{29} =$~536,870,912. 
Above this limit, the problem size does not fit into the DRAM in the measurement environment. 
The total execution time is around $220$--$230$ seconds for $N_\mathrm{tot} \lesssim 10^8$. 
It increases for $N_\mathrm{tot} \gtrsim 10^8$ and reaches $420$ seconds for $N_\mathrm{tot} = 2^{29}$. 
For all values of $N_\mathrm{tot}$, the dominant process is the preparation of the disc components (filled squares), which includes the time required to generate the potential-density pairs and calculate the spherically averaged density profiles, circular velocity, and velocity dispersion in the vertical direction. 
The execution time for this procedure is independent of $N_\mathrm{tot}$ and is $212.0$ seconds on average. 
If there are no disc components, the total execution time is reduced to $10$ seconds for $N_\mathrm{tot} \lesssim 10^7$. 
The time required to prepare the spherical components (filled circles)--which includes setting the radial density profiles, integrating them and calculating the DF using Eddington's formula)--and the time required to write miscellaneous files for further analysis (the open triangles) are also independent of $N_\mathrm{tot}$: they are about $3.4$ and $4.6$ seconds on average, respectively. 
On the other hand, the execution time required to distribute the $N$-body particles (the filled diamonds and solid line) and to write the particle data (the open diamonds and solid line) increase with $N_\mathrm{tot}$, as expected. 
Above $N_\mathrm{tot} \sim 2 \times 10^8$, the writing speed for particle data becomes slow. 
Protocol switching in HDF5 or the bandwidth of the file system (4 WD40EFRXs managed as RAID 5 by mdadm v3.4 and XFS format) might be the reason, but investigating this is beyond the scope of the present study. 
In summary, the measured performance results show that: (1) generating galaxies with disc component(s) takes $4$ minutes for $N_\mathrm{tot} \lesssim 10^8$, (2) generating galaxies without disc components take $10$ seconds for $N_\mathrm{tot} \lesssim 10^7$, and (3) the elapsed time increases for $N_\mathrm{tot} \gtrsim 10^7$. 
Since the most time-consuming procedure in \program{MAGI} is the iterative calculation of the potential-density pairs for the disc components, a more sophisticated preconditioner than ILU(0), or performance optimisations of sparse matrix-vector multiplication would accelerate this part. 

In the default configuration of \program{MAGI}, pseudo random-numbers are generated by \program{SFMT} using the jump function to obtain the benefits of OpenMP thread parallelization. 
Without the jump function, we observe approximately a ten-fold deceleration in the particle-distribution process (the filled diamonds and dotted line in Fig.~\ref{fig:discussion:execution.time}). 
The speed-up rate delivered by the jump function reaches a factor of $10$ for $N_\mathrm{tot} \gtrsim 10^8$, which corresponds to a parallel efficiency of $63$\%. 
GSL also provides a pseudo-random number generator based on the Mersenne Twister of period $2^{19937} - 1$, and it can also be used in \program{MAGI}. 
Since GSL does not provide an equivalent to the jump function, the independence of multiple series of the random numbers it generates is not strictly guaranteed, and OpenMP parallelization must be switched off during the particle-distribution process. 
The performance of the random numbers generator provided in GSL is $4.6$ times slower than \program{SFMT}, and the particle-distribution process requires a $\sim 30$\% longer execution time compared to \program{SFMT} without the jump function. 

The HDF5 container supports on-the-fly file compression with Szip\footnote{\url{https://support.hdfgroup.org/doc\_resource/SZIP/}} 2.1, and the resulting file size of the particle data is $46$\% of the original. 
However, we have not employed Szip compression in the default operation of \program{MAGI}, since it is $10$--$30$ times slower than without Szip compression in most cases. 

\subsection{Strengths and Versatility of MAGI}
\label{subsec:discussion:potential}
Initial conditions for $N$-body simulations are also useful for fitting observed datasets. 
Typically, when fitting an observed dataset, one must generate many mass distribution models and pick an optimal one. 
Three features of \program{MAGI} that makes this procedure very efficient are: (1) \program{MAGI} supports various kinds of density profiles including machine-readable tables; (2) the input parameters adopted in \program{MAGI} are structural ones (total mass and scale length), which helps users to adjust the generated mass-distribution models to reproduce the observed datasets; and (3) the short execution time for \program{MAGI} ($10$ seconds for early-type galaxy models and $4$ minutes for late-type galaxy models) enables users to produce many trial models to fit the observed galaxies. 
Gaia provides six-dimensional phase-space information about the Milky Way's stellar component \citep{GaiaDR1_astrometry}, and direct comparisons between the Gaia data and particle systems generated by \program{MAGI} will be useful for modelling the Milky Way's DF. 

The flexibility of \program{MAGI} also enables one to use idealised simulations to study any kind of physics in detail. 
Most initial-condition generators are limited to astrophysical profiles such as the NFW profile or de Vaucouleurs's law. 
On the other hand, \program{MAGI} accommodates arbitrary spherically symmetric density profiles, so long as the profile is twice differentiable. 
Among other things, this is sometimes useful for testing predictions of analytic models. 
For example, \citet{MichikoshiKokubo2014, MichikoshiKokubo2016} investigated the development of stellar spirals in disc galaxies using linear analysis and local $N$-body simulations, and they estimated the number of spiral arms, assuming a constant ratio of disc mass to the total mass of the galaxy. 
In order to verify their predictions, one would need to carry out global $N$-body simulations with carefully controlled initial conditions. 
The flexibility of \program{MAGI} facilitates the generation of initial conditions for such highly idealised problems. 

Hydrodynamic simulations based on a particle method (e.g., smoothed particle hydrodynamics) are widely employed to investigate in detail the formation and evolution processes of galaxies. 
Extensions to include gaseous components would be meaningful and not difficult, if we assume hydrostatic equilibrium for the gaseous component(s). 

\section{Conclusions}
\label{section:conclusions}
We have developed an initial-condition generator for $N$-body simulations, called \program{MAGI}. 
The implementation relies on DFs to determine the velocity distribution of the target system. 
In order to calculate the DFs, we have exploited Eddington's formula for spherically symmetric components assuming an isotropic velocity distribution, and the Schwarzschild DF with an isothermal profile in the vertical direction for disc components. 
Since \program{MAGI} supports various kinds of volume-density profiles and surface-density profiles including a machine-readable tabular format, \program{MAGI} is very flexible in producing model galaxies. 
One of the strong points of \program{MAGI} is that the software can generate galaxy models including multiple disc components, appropriate for late-type galaxies like the Milky Way. 
Input parameters for each component are limited to the scale length, mass, and other model-specific dimensionless parameters to control the output intuitively. 

We tested the dynamical stability of the generated particle systems using $N$-body simulations. 
The results show that the systems retain the original distribution for at least $1$~Gyr both early-type and late-type galaxy models. 
The execution time required for \program{MAGI} to build a late-type galaxy model is $4$ minutes for $N \lesssim 10^8$. 
Further acceleration can be achieved by improving the algorithm or by performance optimisation of the calculation for the potential-density pair of the disc components. 
The code is provided as open-source software and is publicly available\footnote{\url{https://bitbucket.org/ymiki/magi}}. 

\section*{Acknowledgements}
YM appreciates Masao Mori for providing comments and discussions essential to extend the applicable scope of the code. 
YM has benefited from the feedback given by Takanobu Kirihara, Kazuki Kato, Naohisa Kusu, and Toru Ishikawa on using a beta version of \program{MAGI} which improved the usability of the software. 
We thank Alexander Y. Wagner for checking the manuscript in detail and for his comments that improved the manuscript. 
This research used computational resources in Centre for Computational Sciences, University of Tsukuba. 
The present study was supported by the Japan Science and Technology Agency's (JST) CREST programme entitled ``Research and Development of Unified Environment on Accelerated Computing and Interconnection for Post-Petascale Era''. 
This research was also supported in part by the Grant-in-Aid for Scientific Research (B) by JSPS (15H03638). 




\bibliographystyle{mnras}
\bibliography{ref} 




\appendix
\section{Volume-Density Profiles}
\label{appendix:rho}
Here, we list the volume-density profiles provided by \program{MAGI}. 
In this section, $\rho_0$ is the scale density and $r_\mathrm{s}$ is the scale radius. 
Also, $x$ is defined as $x \equiv r / r_\mathrm{s}$. 
The following three models each have a dense core in their central region. 
For the Plummer sphere \citep{Plummer1911}, the density profile and the first and second derivatives are, respectively, given by 
\begin{align}
  \rho(x) &= \rho_0 \left(1 + x^2\right)^{-5 / 2},
  \\
  \dv{\rho(x)}{x} &= -\frac{5 x}{1 + x^2} \rho(x),
  \\
  \dv[2]{\rho(x)}{x} &= 5 \frac{6 x^2 - 1}{(1 + x^2)^2} \rho(x).
\end{align}
The analytic DF for the Plummer sphere is given by \citet{Aarseth1974} as 
\begin{align}
  f(E) &= \left(\frac{3}{2 \pi G \rho_0 r_\mathrm{s}^2}\right)^5 \frac{\sqrt{2} \rho_0}{7 \pi^2} (-E)^{7 / 2},
  \label{eq:app:Plummer:DF}
  \\
  E &= \frac{v^2}{2} - \frac{G M_\mathrm{tot}}{\sqrt{r^2 + r_\mathrm{s}^2}},
\end{align}
where the total mass $M_\mathrm{tot}$ is $4 \pi \rho_0 r_\mathrm{s}^3 / 3$. 
The density profile and the derivatives for the Burkert sphere \citep{Burkert1995} are 
\begin{align}
  \rho(x) &= \frac{\rho_0}{(1 + x) (1 + x^2)},
  \\
  \dv{\rho(x)}{x} &= - \frac{1 + x (2 + 3 x)}{(1 + x) (1 + x^2)} \rho(x),
  \\
  \dv[2]{\rho(x)}{x} &= 4 x^2 \frac{4 x + 3 \left(1 + x^2\right)}{(1 + x)^2 (1 + x^2)^2} \rho(x).
\end{align}
For the King profile, the comparable results are given in the empirical forms \citep{King1962}
\begin{align}
  \rho(x) &= \rho_0 \left\{\left(1 + x^2\right)^{-1 / 2} - C\right\}^2,
  \quad
  C \equiv \left\{1 + \left(\frac{r_t}{r_\mathrm{s}}\right)^2\right\}^{-1 / 2},
  \\
  \dv{\rho(x)}{x} &= -2 \rho_0 x \left(1 + x^2\right)^{-3 / 2} \left\{\left(1 + x^2\right)^{-1 / 2} - C\right\},
  \\
  \dv[2]{\rho(x)}{x} &= 2 \rho_0 \frac{3 x^2 - 1 + C (1 + x^2)^{1 / 2} (1 - 2 x^2)}{(1 + x^2)^3},
\end{align}
where $r_t$ is the tidal radius. 

The four models below each have a central cusp in their density profiles. 
For the Hernquist model \citep{Hernquist1990}, the volume-density profile and its derivatives are 
\begin{align}
  \rho(x) &= \frac{\rho_0}{x (1 + x)^3},
  \\
  \dv{\rho(x)}{x} &= -\frac{1 + 4 x}{x (1 + x)} \rho(x),
  \\
  \dv[2]{\rho(x)}{x} &= 2 \frac{1 + 5 x (1 + 2 x)}{x^2 (1 + x)^2} \rho(x),
\end{align}
and the analytic DF is given by \citet{Hernquist1990} as 
\begin{align}
  f(E) &= \left(2 \pi G \rho_0 r_\mathrm{s}^2\right)^{-3 / 2} \frac{\rho_0}{2^{5 / 2} \pi^2} \left(1 - q^2\right)^{-5 / 2}
  \notag\\&\hphantom{=} \times
  \left\{3 \arcsin{q} + q \sqrt{1 - q^2} (1 - 2 q^2) (8 q^4 - 8 q^2 - 3)\right\},
  \label{eq:app:Hernquist:DF}
  \\&
  q \equiv \sqrt{\frac{-E}{2 \pi G \rho_0 r_\mathrm{s}^2}},
  \quad
  E = \frac{v^2}{2} - \frac{G M_\mathrm{tot}}{r + r_\mathrm{s}},
\end{align}
where the total mass $M_\mathrm{tot}$ is $2 \pi \rho_0 r_\mathrm{s}^3$. 
For the NFW profile \citep{Navarro1995}, the density profile and the derivatives are 
\begin{align}
  \rho(x) &= \frac{\rho_0}{x (1 + x)^2},
  \\
  \dv{\rho(x)}{x} &= -\frac{1 + 3 x}{x (1 + x)} \rho(x),
  \\
  \dv[2]{\rho(x)}{x} &= 2 \frac{1 + 2 x (2 + 3 x)}{x^2 (1 + x)^2} \rho(x),
\end{align}
while those of the Moore profile \citep{FukushigeMakino1997, Moore1998} are 
\begin{align}
  \rho(x) &= \frac{\rho_0}{x^{3 / 2} (1 + x)^{3 / 2}},
  \\
  \dv{\rho(x)}{x} &= -\frac{3}{2} \frac{1 + 2 x}{x (1 + x)} \rho(x),
  \\
  \dv[2]{\rho(x)}{x} &= \frac{3}{4} \frac{5 + 16 x (1 + x)}{x^2 (1 + x)^2} \rho(x),
\end{align}
and those of the Einasto profile \citep{Einasto1965, Navarro2004, Navarro2010} are 
\begin{align}
  \rho(x) &= \rho_0 e^{2 (1 - x^\alpha) / \alpha},
  \\
  \dv{\rho(x)}{x} &= - 2 x^{\alpha - 1} \rho(x),
  \\
  \dv[2]{\rho(x)}{x} &= 2 x^{\alpha - 2} \left(2 x^\alpha + 1 - \alpha\right) \rho(x),
\end{align}
where $\alpha$ is a parameter that determines the degree of steepness of the transition about the density slope. 

The two remaining models are broken-power-law profiles. 
For the double-power-law model, the so-called ($\alpha$, $\beta$, $\gamma$) model \citep{Hernquist1990, Merritt2006}, the density profile and the derivatives are 
\begin{align}
  \rho(x) &= \rho_0 x^{-\alpha} \left(1 + x^\beta\right)^{(\alpha - \gamma) / \beta},
  \label{eq:methods:eddington:double}
  \\
  \dv{\rho(x)}{x} &= - \frac{\alpha + \gamma x^\beta}{x (1 + x^\beta)} \rho(x),
  \label{eq:methods:eddington:double:drho_dx}
  \\
  \dv[2]{\rho(x)}{x} &= \frac{(\alpha + \gamma x^\beta) \{1 + \alpha + (1 + \beta + \gamma) x^\beta\} - \beta \gamma (1 + x^\beta) x^\beta}{x^2 (1 + x^\beta)^2} \rho(x),
  \label{eq:methods:eddington:double:d2rho_dx2}
\end{align}
where $\beta$ determines the sharpness of the transition from the inner-power-law slope $\alpha$ to the outer-power-law slope $\gamma$. 
The last, the triple-power-law profile, is a natural extension of equation~(\ref{eq:methods:eddington:double}). 
The density profile is defined as 
\begin{align}
  \rho(x) &= \rho_0 x^{-\alpha} \left(1 + x^\beta\right)^{(\alpha - \gamma) / \beta} S(x),
  \label{eq:methods:eddington:triple}
  \\&
  S(x) \equiv \left(1 + y(x)\right)^{(\gamma - \epsilon) / \delta},
  \quad
  y(x) \equiv \left(\frac{r_\mathrm{in}}{r_\mathrm{out}} x\right)^\delta,
\end{align}
where $x = r / r_\mathrm{in}$, $r_\mathrm{in}$ is the inner scale radius and $r_\mathrm{out}$ is the outer scale radius. 
Additional parameters ($r_\mathrm{out}$, $\delta$, $\epsilon$) are introduced to set another density slope at the outskirts of the profile. 
The first and the second derivatives are calculated by using the chain rule with the derivatives of the double-power-law profile , i.e. equations~(\ref{eq:methods:eddington:double:drho_dx}) and (\ref{eq:methods:eddington:double:d2rho_dx2}), and 
\begin{align}
  \dv{S(x)}{x} &= (\gamma - \epsilon) \frac{y(x)}{x \left(1 + y(x)\right)} S(x),
  \\
  \dv[2]{S(x)}{x} &= \frac{\delta - 1 + (\gamma - \epsilon - 1) y(x)}{x \left(1 + y(x)\right)} \dv{S(x)}{x}.
\end{align}

\section{Description of the King Model}
\label{appendix:King}
The DF for the King model (also called the lowered isothermal model) is given by 
\begin{equation}
  f(\mathcal{E}) =
  \left\{\begin{array}{cl}
  \frac{\rho_1}{\left(2 \pi \sigma^2\right)^{3/2}} \left(e^{\mathcal{E}/\sigma^2} - 1\right), & \textrm{  (}\mathcal{E}>0\textrm{)} \\
  0,                                                                                     & \textrm{  (}\mathcal{E}\leq 0\textrm{)}
  \end{array}\right.
  \label{eq:app:King:DF}
\end{equation}
where $\rho_1$ is the scale density, and $\sigma$ is a parameter having the dimension of velocity (but it is not the velocity dispersion). 
Integrating equation~(\ref{eq:app:King:DF}) over velocity space gives the density profile: 
\begin{equation}
  \rho = \rho_1 \left\{e^{W} \mathrm{erf}\left(\sqrt{W}\right) - \sqrt{\frac{4 W}{\pi}} \left(1 + \frac{2 W}{3}\right)\right\}, 
\end{equation}
where $W \equiv \Psi / \sigma^2$ is the dimensionless potential. 
It is determined by numerically solving Poisson's equation 
\begin{equation}
  \dv{x}\left(x^2 \dv{W}{x}\right)
  = - \frac{9 \rho_1 x^2}{\rho_0} \left\{e^{W} \mathrm{erf}\left(\sqrt{W}\right) - \frac{2 \sqrt{W}}{3 \sqrt{\pi}} \left(3 + 2 W\right)\right\},
  \label{eq:app:King:Poisson.eq}
\end{equation}
where $x$ denotes the radius normalized by the King radius $r_0$. 
When solving equation~(\ref{eq:app:King:Poisson.eq}), two boundary conditions are necessary to specify the solution: the non-dimensional King parameter $W_0$ at the centre and the requirement of a central core ($\dv*{W}{x} = 0$ at the centre). 
Using the central density $\rho_0$, the King radius $r_0$ is given by 
\begin{equation}
  r_0 = \sqrt{\frac{9 \sigma^2}{4 \pi G \rho_0}}.
\end{equation}

The first and second derivatives of the density profile can then be expressed as 
\begin{equation}
  \dv{\rho}{r} = \frac{\rho_1}{r_0} \left[e^{W} \mathrm{erf}\left(\sqrt{W}\right) - \frac{2 \sqrt{W}}{\sqrt{\pi}}\right] \dv{W}{x},
\end{equation}
\begin{equation}
  \dv[2]{\rho}{r} = \frac{\rho_1}{{r_0}^2} \left[\left\{e^{W} \mathrm{erf}\left(\sqrt{W}\right) - \frac{2 \sqrt{W}}{\sqrt{\pi}}\right\} \dv[2]{W}{x} + e^{W} \mathrm{erf}\left(\sqrt{W}\right) \left(\dv{W}{x}\right)^2\right]. 
\end{equation}
Here, we already have $\dv*{W}{x}$ from the solution of equation~(\ref{eq:app:King:Poisson.eq}), and $\dv*[2]{W}{x}$ is obtained by solving the differential equation 
\begin{equation}
  \dv[2]{W}{x} = - \frac{9 \rho_1}{\rho_0} \left[e^{W} \mathrm{erf}\left(\sqrt{W}\right) - \frac{2 \sqrt{W}}{3 \sqrt{\pi}} \left(3 + 2 W\right)\right] - \frac{2}{x} \dv{W}{x}. 
\end{equation}

\section{Surface-Density Profiles}
\label{appendix:Sigma}
In this section, we list the surface-density profiles provided by \program{MAGI}. 
Here, $X$ is defined as $X \equiv R / R_\mathrm{s}$, where $R_\mathrm{s}$ is the scale length. 
The surface-density profile for the S\'ersic model \citep{Sersic1963} is given by 
\begin{equation}
  \Sigma(X) = \Sigma_0 \exp{\left(-b_n X^{1 / n}\right)},
\end{equation}
where $n$ is the S\'ersic index and $b_n$ is a dimensionless scale factor. 
The scale factor $b_n$ is internally calculated using an asymptotic formula by \citet{CiottiBertin1999} that works well, at least in the range $1 \leq n \leq 10$. 
S\'ersic profiles with $n = 1$ and $4$ correspond to an exponential profile and de Vaucouleurs's law \citep{deVaucouleurs1948}, respectively. 

\section{Radial Profiles of Disc Components}
\label{appendix:Disc}
\begin{figure}
  \includegraphics[width=\columnwidth, clip]{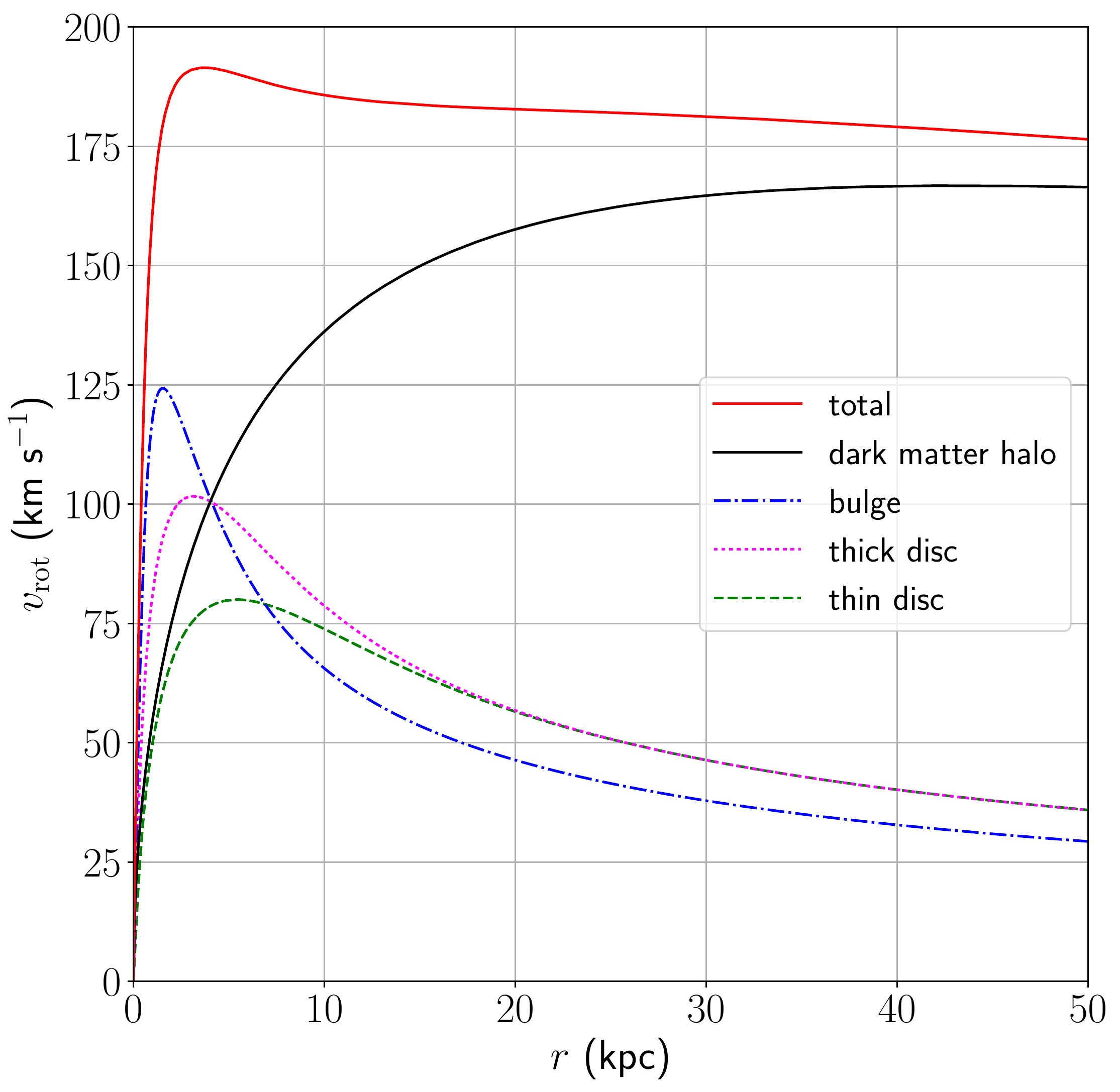}
  \caption{
    Rotation curve for the late-type galaxy model of \S\ref{subsec:results:ltg}. 
    The black solid, blue dot-dashed, magenta dotted, and green dashed curves show the contributions from the dark-matter halo, bulge, thick disc, and thin disc, respectively. 
    The solid red curve is the sum of all components. 
  }
  \label{fig:app:disc:vrot}
\end{figure}
\begin{figure}
  \includegraphics[width=\columnwidth, clip]{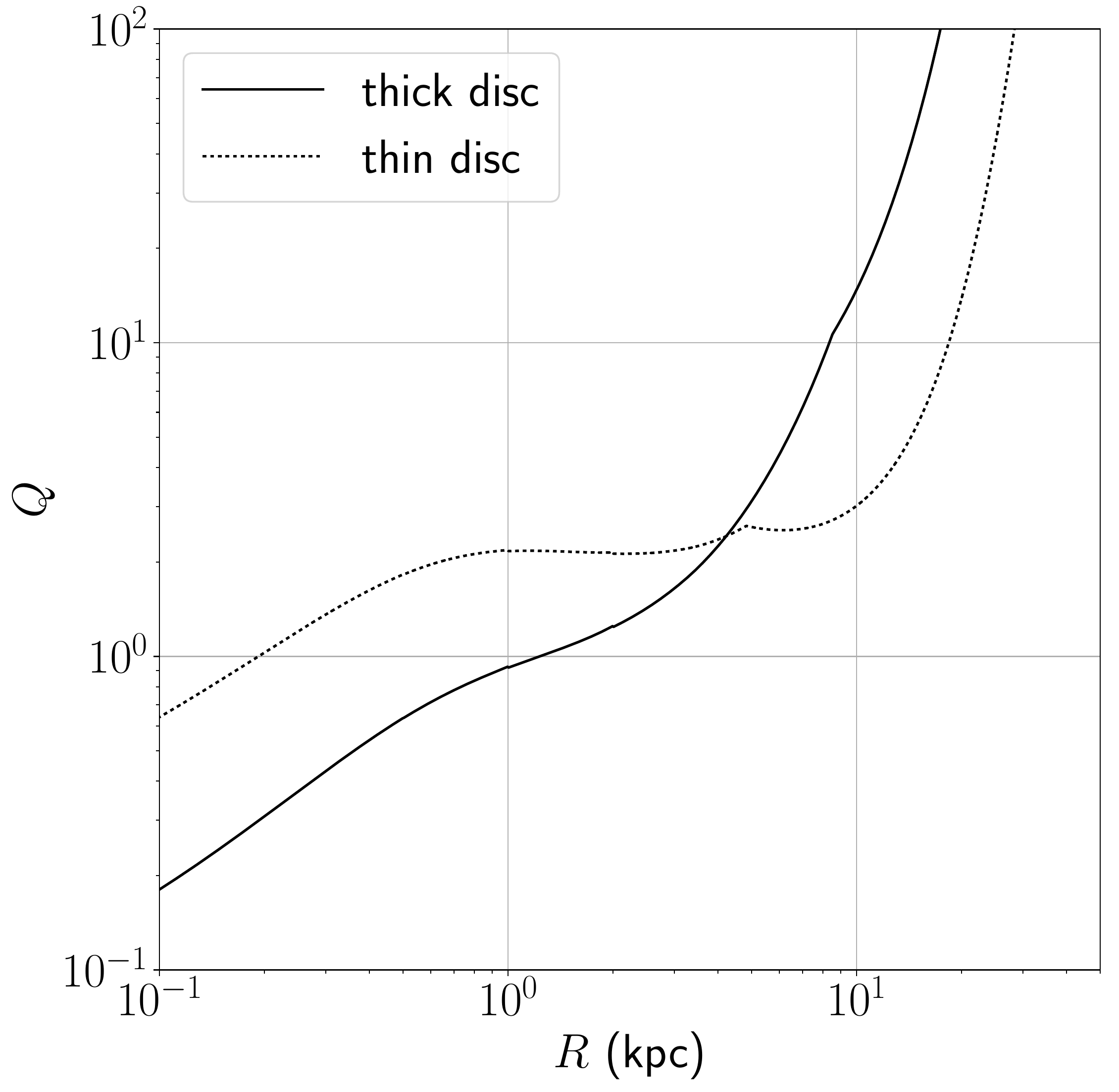}
  \caption{
    Radial profiles of Toomre's $Q$-value for the disc components. 
  }
  \label{fig:app:disc:Q}
\end{figure}
\begin{figure}
  \includegraphics[width=\columnwidth, clip]{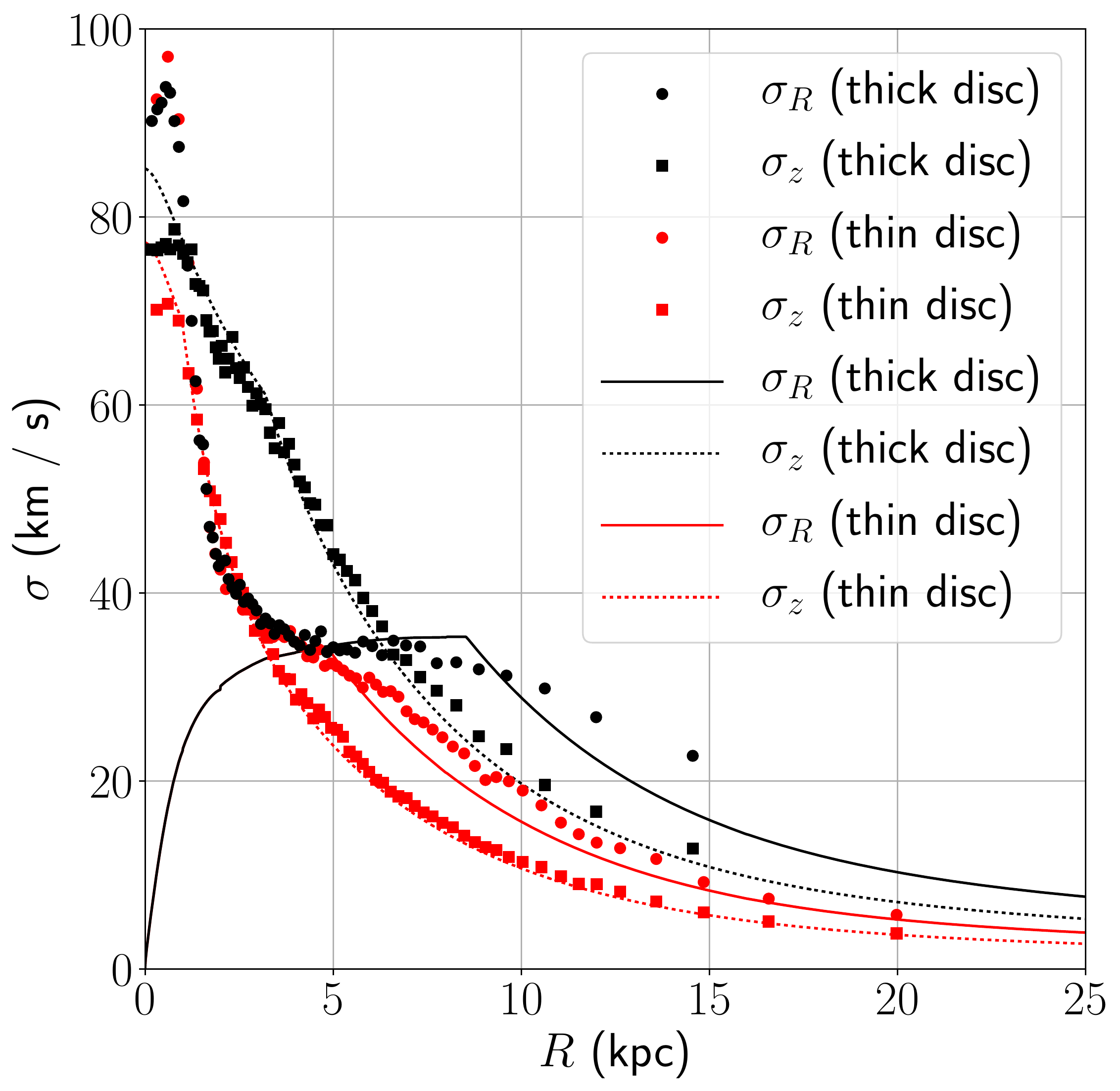}
  \caption{
    Radial profiles of the velocity dispersion for the disc components. 
    Symbols and curves show the profile at $t = 1$~Gyr and the input profile, respectively. 
    The black symbols and lines show the thick-disc component, while the red ones correspond to the thin-disc component. 
    The circles and solid curves represent the velocity dispersion in the $R$ direction, while the squares and dotted curves correspond to the velocity dispersion in the $z$ direction. 
  }
  \label{fig:app:disc:sigma}
\end{figure}
In this section, we show radial profiles of the late-type galaxy model investigated in \S\ref{subsec:results:ltg}. 
Figure~\ref{fig:app:disc:vrot} shows the rotation curve of the model galaxy. 
Figure~\ref{fig:app:disc:Q} exhibits the radial profiles of Toomre's $Q$-value for the disc components. 
Since the velocity dispersion of the disc components is capped in the central region to ensure that the epicycle approximation remains valid, the $Q$-value decreases toward the centre and becomes smaller than unity. 
Figure~\ref{fig:app:disc:sigma} compares the radial profiles of the velocity dispersion for the disc components at $t=1$~Gyr (symbols) with the input radial profiles (curves). 
The figure shows that the given profiles of $\sigma_z$ are stable over the entire domain after an integration time of $1$~Gyr. 
On the other hand, the radial profiles of $\sigma_R$ evolve, especially in the central region. 
This indicates that the epicycle approximation is not satisfied in the central regions of late-type galaxies. 
The $\sigma_R$-profile of the thin-disc component has better stability in other domains, which suggests that the thin disc is more consistent with the epicycle approximation than the thick-disc component. 


\bsp	
\label{lastpage}
\end{document}